%% file: geomInteg.tex
\documentclass[
  twoside,
]{interact}
\newcommand{\DFmode}{final}
\usepackage[\DFmode]{ifdraft}
\ifdraft{\overfullrule=5pt}{}

\usepackage[svgnames]{xcolor}

\usepackage{layout}
\usepackage[\DFmode,color,notcite,notref]{showkeys}

\definecolor{labelkey}{cmyk}{0.00, 0.420, 0.210, 0.050}
\definecolor{refkey}{cmyk}{0.240, 0.020, 0.000, 0.000}
\usepackage[sc,osf,slantedGreek]{mathpazo}
\usepackage[T1]{fontenc}
\IfFileExists{euscript.sty}{%
  \usepackage[mathcal]{euscript}}{}
\usepackage{microtype}
\usepackage{textcase}
\DeclareTextFontCommand{\textsmallcaps}{\scshape}

\newcommand{\smallcapsspacing}[1]{\textls[50]{##1}}

\newcommand{\smallcaps}[1]%
  {\smallcapsspacing{\scshape\MakeTextLowercase{##1}}}
\renewcommand{\textsc}[1]{\smallcapsspacing{\textsmallcaps{##1}}}
\usepackage{relsize}

\usepackage{fancyhdr}
\usepackage{tocloft}
\cftsetrmarg{21pt}
\cftsetpnumwidth{15pt}

\usepackage{array}
\newcolumntype{C}{>{$}c<{$}}
\newcolumntype{L}{>{$}l<{$}}
\newcolumntype{R}{>{$}r<{$}}
\usepackage{booktabs}
\usepackage{colortbl}
\usepackage{multirow}
\usepackage{bigstrut}
\usepackage{tabularx}

\usepackage{mathtools}

\usepackage{epstopdf}
\usepackage[caption=false]{subfig}
\graphicspath{{./figures/}{./figures/_new_figures_/}%
             }

\PassOptionsToPackage{hyphens}{url}
\usepackage[final=true,colorlinks=true]{hyperref}
\definecolor{hrBlue}{cmyk}{1.00, 0.25, 0.25, 0.00}
\definecolor{hrRed}{cmyk}{0.25, 1.00, 1.00, 0.00}
\definecolor{hrGreen}{cmyk}{1.00, 0.25, 1.00, 0.00}
\hypersetup{
  pdfborder = {0 0 0},
  bookmarksdepth = section,
  citecolor = hrGreen,
  linkcolor = hrRed,
  urlcolor  = hrBlue
}
\usepackage[numbers,sort&compress]{natbib}
\bibpunct[, ]{[}{]}{,}{n}{,}{,}

\usepackage{bibentry}

\usepackage{notes2bib}

\theoremstyle{plain}
\newtheorem{theorem}{Theorem}[section]

\theoremstyle{definition}

\theoremstyle{remark}

\usepackage{siunitx}

\usepackage{fixme}
\ifdraft{\fxsetup{status=draft}}%
        {\fxsetup{status=final}}
\fxsetup{theme=color,mode=multiuser}
\fxsetup{layout=margin,innerlayout={layout=inline}}

\fxsetface{inline}{\itshape\small}
\fxsetface{margin}{\itshape\footnotesize}
\FXRegisterAuthor{ajd}{aajd}{AJD}
\FXRegisterAuthor{dta}{adta}{DTA}
\definecolor{fxtarget}{rgb}{0.7725, 0.2275, 0.2275}

\usepackage{braket}

\usepackage{overpic}

\usepackage[boldvec]{dta}
\usepackage{liecolon}

\DeclareMathOperator{\ispm}{ispm}
\DeclareMathOperator{\ISpM}{ISpM}

\newcommand*{\U}[1]{\ensuremath{\mathrm{U}(#1)}}
\newcommand*{\SpR}[1]{\ensuremath{\mathrm{Sp}(#1,\R)}}

\newcommand*{\R}[1][\mspace{-2mu}]{\ensuremath{\mathbb{R}^{\mspace{2mu}#1}}}

\newcommand*{\Gl}[1][l]{G^{(#1)}}
\newcommand*{\Ql}[1][l]{Q^{(#1)}}
\newcommand*{\Graml}[1][l]{\Gamma(#1)}

\newcommand*{\pcan}{\ensuremath{\V{p}^\text{can}}}
\newcommand*{\pmech}{\ensuremath{\V{p}^\text{mech}}}

\newcommand*{\prmech}{\ensuremath{p^\text{mech}}}
\newcommand*{\maxm}{\text{maxm}}
\newcommand*{\trunc}{\text{trunc}}
\newcommand*{\Mtr}[1]{\ensuremath{\Map{M}^{\text{tr#1}}}}
\newcommand*{\Ntr}[1]{\ensuremath{\Map{N}^{\hmhsp\text{tr#1}}}}
\newcommand*{\Mpsc}[1]{\ensuremath{\Map{M}^{\text{psc#1}}}}
\newcommand*{\Npsc}[1]{\ensuremath{\Map{N}^{\hmhsp\text{psc#1}}}}
\newcommand*{\Mcr}[1][]{\ensuremath{\Map{M}^{\text{cr#1}}}}
\newcommand*{\Ncr}[1][]{\ensuremath{\Map{N}^{\hmhsp\text{cr#1}}}}
\newcommand*{\Mrtcr}[1][]{\ensuremath{\Map{M}^{\text{rtcr#1}}}}
\newcommand*{\Nrtcr}[1][]{\ensuremath{\Map{N}^{\hmhsp\text{rtcr#1}}}}
\newcommand*{\Mrtscr}[1][]{\ensuremath{\Map{M}^{\text{rtscr#1}}}}
\newcommand*{\Nrtscr}[1][]{\ensuremath{\Map{N}^{\hmhsp\text{rtscr#1}}}}

\input{ushyphex}
\articletype{Book Chapter}
\title{Structure-preserving techniques in accelerator physics}
\author{%
  \name{Dan T. Abell\textsuperscript{a}
        and
        Alex J. Dragt\textsuperscript{b}}%
  \affil{\textsuperscript{a}RadiaSoft LLC, Boulder, Colorado, USA}%
  \affil{\textsuperscript{b}University of Maryland Physics Department
                            (Professor Emeritus)}%
         \thanks{CONTACT A.\,J.\ Dragt
                 Email: \texttt{\href{mailto:dragtnb@comcast.net}%
                                            {dragtnb@comcast.net}}}%
\vspace{-\baselineskip}
}
\renewcommand{\today}{31 October 2022}

\begin{document}
\maketitle
\thispagestyle{empty}

\nobibliography*

\begin{abstract}
To a very good approximation, particularly for hadron machines,
charged-particle trajectories in accelerators obey Hamiltonian mechanics.
During routine storage times of eight hours or more,
such particles execute some
\num{e8} revolutions about the machine,
\num{e10} oscillations about the design orbit,
and \num{e13} passages through various bending and focusing elements.
Prior to building, or modifying, such a machine,
we seek to identify accurately the long-term behavior and stability
of particle orbits over such large numbers of interactions.
This demanding computational effort does not yield easily
to traditional methods of symplectic numerical integration,
including both explicit Yoshida-type
and implicit Runge-Kutta or Gaussian methods.
As an alternative, one may compute an approximate one-turn map
and then iterate that map.
We describe some of the essential considerations and techniques
for constructing such maps to high order
and for realistic magnetic field models.
Particular attention is given to preserving
the symplectic condition characteristic of Hamiltonian mechanics.
\end{abstract}


\begingroup
  \renewcommand*{\addvspace}[1]{\vskip4.4pt}
  \tableofcontents
\endgroup
\thispagestyle{empty}

\ifdraft{%
  \newpage
  \listoffixmes
}{}

\setcounter{footnote}{0}

\ifdraft{%
  \newpage
}{}
\pagestyle{fancy}
%
\input{Intro}
\input{LandH}

\input{TransferMaps}

\input{LieTools}

\input{SympJets}

\input{GenFns}

\input{Cremona}

\input{Conclude}

\input{Ackn}


\ifdraft{\addcontentsline{toc}{section}{References}}{}
\addcontentsline{toc}{section}{References}
\bibliographystyle{ieeetr}
\bibliography{geomInteg}







\end{document}

%% file: Intro.tex

\section{Introduction}
\label{sec:intro}

Particle accelerators consist of arrays of magnets and radio-frequency (rf)
cavities.  Their purpose is to produce intense beams of high-energy charged
particles including electrons, positrons (the antimatter counterpart
of electrons), protons, antiprotons, and various ions.  The magnets provide
magnetic fields that bend, focus, and exert various nonlinear effects on
beam particle orbits; and the rf cavities provide electric fields
that accelerate and longitudinally bunch particles.

Successful accelerator design requires the accurate calculation/simulation
of particle orbits over long periods of time.  The most computationally
challenging are orbits in \emph{storage rings}, essentially circular machines
in which particles continually circulate for long periods of time.
Electron (or positron) storage rings are used to produce intense high-energy
X-rays.  Electron (or positron) storage rings, as well as proton
(or antiproton or ion) storage rings, are used in pairs to produce colliders.
In a collider, the beam from one ring collides head-on with the
counter-circulating beam in a second ring.%
  \footnote{But, even with the highest achievable beam densities,
            beam-beam collisions are sufficiently rare that the stored
            beams are only partially depleted over the storage time.
            However, the colliding beams can have significant dynamical
            (both linear and nonlinear) effects on each other, a complication
            that must be understood/managed but lies  beyond the scope of
            this chapter.}
For example, the Large Hadron Collider (LHC) in CERN
(near Geneva, Switzerland), which collides protons on protons,
has a circumference of \SI{27}{km}, and each ring has
some \num{19200}~elements
(magnets, rf cavities, and intervening drift spaces).
Protons moving at essentially the speed of light are stored for about
\SI{8}{hours}, and during this time make
approximately \num{ 4e 8}~turns around the ring,
approximately \num{ 8e12}~element passages, and
approximately \num{20e 9}~(betatron) oscillations about the design orbit.
Following this number of oscillations is comparable to following
the earth's orbit about the sun from the time of the Big Bang.
While the number of betatron oscillations in electron (or positron)
storage rings is comparable, they need not be followed for as long
because those oscillations are damped by the energy loss associated
with X-ray emission.
Correspondingly, electron/positron storage-ring orbits are less
computationally challenging than proton/antiproton/ion storage-ring orbits.
This chapter is devoted to the most challenging problem of
calculating/simulating particle orbits over long periods of time
in proton (or antiproton or ion) storage rings.

%% file: LandH.tex

\section{Lagrangians and Hamiltonians}
\label{sec:LandH}

In Cartesian coordinates, the relativistic Lagrangian $L$ for the motion
of a particle of mass $m$ and charge $q$ in an electromagnetic field
is given by the expression
\begin{equation}
  \label{eq:lagrange}
  L(\V{r}, \V{v}, t) = -m c^2(1 - v^2/c^2)^{1/2}
    - q \psi(\V{r}, t) + q \V{v} \cdot \V{A}(\V{r}, t).
\end{equation}
Here $\V{r}$ is the particle position at time $t$, $\V{v} = \dt{\V{r}}$
is the particle velocity, and $c$ is the speed of light.
The quantities $\psi$ and $\V{A}$ are the scalar and vector potentials
defined in such a way that the electromagnetic fields $\V{E}$ and $\V{B}$
are given by the standard relations
\begin{subequations}\label{eq:fields}
\vspace{-\baselineskip}
\begin{align}
  \label{eq:magB}
  \V{B} &= \curl\V{A}, \\
  \label{eq:elecE}
  \V{E} &= -\grad\psi - \ptdd{\V{A}}{t}.
\end{align}
\end{subequations}
This formulation ignores spin, radiation reaction (X-ray emission,
also referred to as synchrotron radiation), and quantum effects%
  ~\bibnote[LM:init]{%
      Much of the background material for this chapter is most easily
      found on the Web in a draft book:
      A.\,J. Dragt, \emph{Lie Methods for Nonlinear Dynamics with
      applications to Accelerator Physics},
      \urlprefix\url{http://www.physics.umd.edu/dsat/dsatliemethods.html}.
      It, in turn, provides numerous additional references.
      In subsequent citations it will be referred to as \emph{LM}.
      For a discussion of Lagrangians and Hamiltonians for
      charged-particle motion in electromagnetic fields,
      see \emph{LM}, Sections~1.5--1.7.%
}.
These effects may be important over long times for lighter particles
such as electrons and positrons, but they are significantly less important
for heavier particles such as protons and antiprotons and ions.

For the Lagrangian \eqref{lagrange} the \emph{canonical} momentum in
Cartesian coordinates is given by the equation
\begin{equation}
  \label{eq:p.can}
  \pcan = \ptdd{L}{\V{v}} = m \V{v} / (1 - v^2 / c^2)^{1/2} + q\V{A}.
\end{equation}
Here we use the superscript \emph{can} to emphasize that \eqref{p.can}
defines the \emph{canonical} momentum.  Note that the first term in
\eqref{p.can} is just the relativistic \emph{mechanical} momentum,
\begin{equation}
  \label{eq:p.mech}
  \pmech = m \V{v} / (1-v^2/c^2)^{1/2} = \gamma m\V{v},
\end{equation}
where $\gamma$ is the standard relativistic factor 
\begin{equation}
  \label{eq:gamma}
  \gamma = 1 / (1 - v^2/c^2)^{1/2}.
\end{equation}
Consequently, the relation \eqref{p.can} may also be written in the forms
\begin{equation}
  \label{eq:pc.pm}
  \pcan = \pmech +  q\V{A}
  \quad\text{and}\quad
  \pmech = \pcan -  q\V{A}.
\end{equation}
Upon implementing the standard procedure that relates Lagrangians and
Hamiltonians, one finds that the Hamiltonian $H$ associated with the
Lagrangian $L$ specified by \eqref{lagrange} is given by the expression
\begin{equation}
  \label{eq:ham}
  H = [m^2 c^4 + c^2 (\pcan - q \V{A}) \cdot (\pcan - q \V{A})]^{1/2}
        + q \psi
    = [m^2 c^4 + c^2 (\pcan - q \V{A})^2]^{1/2} + q \psi.
\end{equation}

In the usual Hamiltonian formulation (as in the usual Lagrangian formulation)
the time $t$ plays the distinguished role of an \emph{independent} variable,
and all the coordinates $q$ and momenta $p$ are \emph{dependent} variables.%
  \footnote{We are embarrassed by the custom of also using the symbol
            $q$ to denote the \emph{charge} of the particle in question.}
That is, the canonical variables are viewed as functions $q(t)$, $p(t)$
of the independent variable $t$.  In some cases, it is more convenient to
take some \emph{coordinate} to be the independent variable rather than
the time, in which case the time becomes a dependent variable.
So doing may facilitate the use of \emph{transfer maps}, as described in
the next section.  For example, consider the passage of a collection
of particles through a rectangular-shaped beam-line element such as
a magnet or an rf cavity.  In such a situation, particles with different
initial conditions will require different times to pass through
the beam-line element.
If the quantities of interest are primarily the locations and momenta
of the particles as they leave the exit face of the beam-line element,
then it would clearly be more convenient to use for an independent
variable a coordinate that measures the progress of a particle through
the beam-line element.
With such a choice, the relation between entering coordinates and momenta
and exiting coordinates and momenta could be treated as a transfer map.
Remarkably, this goal can be achieved within a Hamiltonian framework%
  ~\bibnote[LM:S.1.6]{\emph{LM}, Section~1.6.}.

\begin{theorem}
Suppose $H(q,p,t)$ is a Hamiltonian for a system having $n$ degrees of
freedom.  Suppose further that $\dot{q}_1 = \ptdd{H}{p_1} \neq 0$
for some interval of time~$T$ in some region~$R$ of the phase space
described by the $2n$ variables $(q_1,\dotsc,q_n)$ and $(p_1,\dotsc,p_n)$.
Then, in this region and time interval, $q_1$ can be introduced as
an independent variable in place of the time $t$.  Moreover, the equations
of motion with $q_1$ as an independent variable can be obtained from
a Hamiltonian that will be called $K$.  To construct $K$, define a quantity
$p_t$ by the rule
\begin{equation}
  \label{eq:pt.def}
  p_t = - H(q,p,t).
\end{equation}
Suppose that this relation is solved for $p_1$ to give a relation of the form
\begin{equation}
  \label{eq:p1}
  p_1 = -K(t,q_2,\dotsc,q_n; p_t,p_2,\dotsc,p_n; q_1).
\end{equation}
Such an inversion is possible according to the inverse function theorem
because $\ptdd{H}{p_1} \neq 0$ by assumption.  Then, as the notation is
intended to suggest, $K$ is the desired new Hamiltonian.
In this formulation, $t$ is treated as a coordinate like the remaining
$q_2$ \dots\ $q_n$, and $p_t$ is its conjugate momentum.
\end{theorem}

As an example, let us use this construction to find the Hamiltonian $K$
corresponding to the Hamiltonian $H$ given by \eqref{ham} when
the $z$ coordinate is taken to be the independent variable.
Assume that  $\dot{z} > 0$ for the trajectories in question.
Then one finds the result
\begin{equation}
  \label{eq:K}
  K = -[(p_t + q\psi)^2 / c^2 - m^2 c^2
          - (p_x - qA_x)^2 - (p_y - qA_y)^2]^{1/2} - qA_z.
\end{equation}
Here the quantities $p_x$ and $p_y$ denote \emph{canonical} momenta.
Note that, according to \eqref{pt.def}, $p_t$ is usually negative.
For the example at hand, one finds that
\begin{equation}
  \label{eq:pt}
  p_t = -[m^2 c^4 + c^2 (\pmech\cdot\pmech)]^{1/2} - q \psi
      = - \gamma mc^2 - q \psi.
\end{equation}

There is yet another Hamiltonian formulation that is of interest.
In the spirit of relativity, and following the insight of
Hermann Minkowski (1864--1909), it is reasonable to try to treat
space and time on a similar footing%
  ~\bibnote[LM:S.1.6e]{\emph{LM}, Section~1.6 and Exercise~1.6.7}.
Let us review some of the mathematical machinery of Special Relativity.
Suppose the world-line
of a particle through space-time is parameterized in terms of some
parameter $\tau$ by specifying four functions $x^\mu(\tau)$ that,
taken together, form a 4-vector with four contravariant components $x^\mu$.
We adopt the convention that the first three components of $x^\mu$ are
the spatial coordinates of the particle, and the fourth (with a factor
of $c$) is its temporal coordinate.
Specifically (for $\mu = 1, 2, 3, 4$ and with $x^4 = ct$) we write
\begin{equation}
  \label{eq:x.mu}
  x^\mu = (x, y, z, ct) = (\V{r}, ct).
\end{equation}

In addition, let $(x^\prime)^\mu$ denote the four derivatives defined by
the equations
\begin{equation}
  \label{eq:dx.mu}
  (x^\prime)^\mu = \dd{x^\mu}{\tau}.
\end{equation}
Under the assumption that the parameterization is unchanged by a Lorentz
transformation, $(x^\prime)^\mu$ is evidently also a 4-vector,
which will be called the 4-velocity.  The 3-velocity, $\V{v}$, of a particle
is given by the ratio $\V{v} = (\dd{\V{r}}{\tau}) / (\dd{t}{\tau})$.
Since the speed of a massive particle must be less than $c$,
$||\V{v}|| < c$, it follows that (for physical particles) the 4-velocity
must satisfy the condition
\begin{equation}
  \label{eq:xp.sq.gt0}
  x^\prime \cdot x^\prime
    = (x^\prime)^\mu (x^\prime)^\nu g_{\mu\nu} > 0.
\end{equation}
Here $g_{\mu\nu}$ denotes the metric tensor, and we have employed the usual
Einstein convention that repeated indices are to be summed over.
In Cartesian coordinates and for flat space-time, only the diagonal entries
of $g$ are nonzero, and we take them to have the values
\begin{equation}
  \label{eq:g.munu}
  g_{11} = g_{22} = g_{33} = - 1,\; g_{44} = 1.
\end{equation}
That is, the space-time interval $\ud{s}$ is taken to be given by
the relation
\begin{equation}
  \label{eq:ds.sq}
  \ud{s}^2 = g_{\mu\nu} \ud{x}^\mu \ud{x}^\nu
    = c^2 \ud{t}^2 - (\ud\V{r})^2.
\end{equation}
We remark that the notation $\ud{s}^2$ appearing in \eqref{ds.sq},
although universally employed, can be misleading since, depending on
circumstances, $\ud{s}^2$ can be negative, zero, or positive, and is
therefore not necessarily the square of anything.
But note that $\ud{s}^2 > 0$ for time-like displacements.
Space-time endowed with the metric \eqref{g.munu} is sometimes called
\emph{Minkowski} space.

The metric tensor can be used to raise and lower indices.
For example, there are the relations
\begin{equation}
  \label{eq:down.up}
  x_\mu = g_{\mu\nu} x^\nu.
\end{equation}
In particular, $x_\mu$ has the entries
\begin{equation}
  \label{eq:x_mu}
   x_\mu = (-x, -y, -z, ct) = (-\V{r}, ct).
\end{equation}
Finally, we define a 4-potential $A^\mu$ with entries
\begin{equation}
  \label{eq:A.mu}
  A^\mu = (A_x, A_y, A_z, \psi/c) = (\V{A}, \psi/c).
\end{equation}

We are now ready to employ some of this mathematical machinery.
Consider the \emph{relativistic} Lagrangian $L_R$ defined by the relation
\begin{equation}
  \label{eq:L.rel}
  L_R = \tfrac{1}{2}mc\, (x^\prime)^\mu (x^\prime)^\nu g_{\mu \nu}
          + q\, (x^\prime)^\mu A^\nu g_{\mu \nu}.
\end{equation}
It has the pleasing property that it is algebraically simple and treats
space and time on a similar footing.  In particular, $L_R$ is evidently
a Lorentz scalar.  That is, it is invariant under Lorentz transformations.%
  \footnote{The quantity $(x^\prime)^\mu A^\nu g_{\mu\nu}$ is a scalar
      under Lorentz transformations provided the 4-potential $A^\nu$
      actually transforms as a 4-vector.
      See~\bibnote[LM:S.1.6fn]{\emph{LM}, footnotes in Exercise~1.6.7.}
      for a discussion of the contrary case.}

The \emph{canonical} momentum $p_\mu$ is given by the relation
\begin{subequations}\label{eq:p_mu}
\begin{equation}
  \label{eq:p_mu.a}
  p_\mu = \ptdd{L_R}{(x^\prime)^\mu} = mc\,(x^\prime)_\mu + qA_\mu,
\end{equation}
which can also be written in the form
\begin{equation}
  \label{eq:p_mu.b}
  p_\mu = \prmech_\mu + qA_\mu,
\end{equation}
where the \emph{mechanical} momentum is given by
\begin{equation}
  \label{eq:p_mu.mech}
  \prmech_\mu = mc\,(x^\prime)_\mu.
\end{equation}
\end{subequations}
According to \eqref{p_mu.mech}, the mechanical momentum transforms like
a 4-vector under Lorentz transformations because $(x^\prime)_\mu$ transforms
like a 4-vector.  From \eqref{p_mu.b} we see that the canonical momentum
also transforms like a 4-vector to the extent that the 4-potential does so.%
  \footnote{Again see~\bibnotemark[LM:S.1.6fn] for a discussion
            of the contrary case.}

Again implementing the standard procedure that relates Lagrangians and
Hamiltonians, one finds that the relativistic Hamiltonian $H_R$ associated
with the Lagrangian $L_R$ specified by \eqref{L.rel} is given by
the expression
\begin{eqnarray}
  \label{eq:H.rel}
  H_R &=& \tfrac{1}{2} mc\, (x^\prime)^\mu (x^\prime)^\nu g_{\mu\nu}
       = \tfrac{1}{2mc} (p^\mu - qA^\mu) (p^\nu - qA^\nu) g_{\mu\nu}
       \nonumber\\
      &=& \tfrac{1}{2mc} (p^\mu - qA^\mu) (p_\mu - qA_\mu).
\end{eqnarray}
Observe that $H_R$, like $L_R$, is Lorentz invariant.
Note also that the phase space associated with world-lines is
eight-dimensional with canonical coordinates
$x^\mu$ and $p_\nu$.

Let us see what can be said about the phase-space trajectories generated
by $H_R$.  Evidently $H_R$, as given by \eqref{H.rel}, does not depend
explicitly on $\tau$,
\begin{equation}
  \label{eq:dLR.0}
  \ptdd{H_R}{\tau} = 0.
\end{equation}
It follows that $H_R$ is a constant (and integral) of motion.
Moreover, from \eqref{H.rel}, we see that the quantity
$\ud{s}^2/(\ud{\tau})^2 $ defined by
\begin{subequations}\label{eq:d2s}
\begin{equation}
  \label{eq:d2s.dt2}
  \ud{s}^2 / (\ud{\tau})^2 = g_{\mu\nu} (x^\prime)^\mu (x^\prime)^\nu
    = (x^\prime)^\mu (x^\prime)_\mu
    = x^\prime \cdot x^\prime
\end{equation}
is a constant (and integral) of motion.
\begin{equation}
  \label{eq:d2s.dt2.c}
  \ud{s}^2/(\ud\tau)^2 = \text{constant}.
\end{equation}
\end{subequations}
Suppose we restrict our attention to those solutions
that satisfy the relation
\begin{equation}
  \label{eq:xp.sq}
  x^\prime \cdot x^\prime = 1.
\end{equation}
From \eqref{p_mu.mech} and \eqref{xp.sq} we see that for these solutions
$(\prmech)^\mu$ satisfies the \emph{mass-shell condition}
\begin{equation}
  \label{eq:m.shell}
  \prmech_\mu (\prmech)^\mu = (\prmech)\cdot(\prmech) = m^2c^2.
\end{equation}
From \eqref{H.rel} and \eqref{xp.sq}, we find that for
these solutions $H_R$ has the value
\begin{equation}
  \label{eq:H.rel.l}
  H_R = (mc/2).
\end{equation}
Moreover, for those solutions that satisfy \eqref{xp.sq}, we have
the result $\ud{s}^2 > 0$ and may therefore select, in accord with
\eqref{ds.sq}, \eqref{d2s.dt2}, and \eqref{xp.sq}, a parameterization
such that
\begin{equation}
  \label{eq:ds.dt.1}
  \dd{s}{\tau} = 1.
\end{equation}

We have introduced three Hamiltonians, namely $H$, $K$, and $H_R$.
It can be shown that they all describe the same physics, and in this sense
are equivalent.%
  \footnote{When using $H_R$, we employ only solutions that
    obey \eqref{xp.sq}.}
Which is to be employed depends on context.
In what follows, we will use the Hamiltonian $K$ with
some reference to the Hamiltonian $H_R$.

%% file: TransferMaps.tex

\section{Transfer Maps, the Symplectic Condition, and Symplectic Integrators}
\label{sec:transfer.maps}

To proceed, it is convenient to introduce
some terminology and definitions.
Suppose we are working with a $2n$-dimensional phase space.
Let the symbol $z$ denote the collection of canonical phase-space variables
arranged in the form
\begin{equation}
  \label{eq:z.def}
  z = (z_1, z_2, \dotsc, z_{2n})
    = (q_1, \dotsc, q_n; p_1, \dotsc, p_n).
\end{equation}
Let $f(z,t)$ and $g(z,t)$ be any two (possibly ``time-dependent'')
functions of $z$.%
  \footnote{Here, by the ``time'', we mean whatever has been selected
    to be the independent variable.  Note also that the symbol $z$ now
    no longer refers to the third component of $\V{r}$, but rather
    to the collection of phase-space variables.}
Define their \emph{Poisson bracket}, $[f,g]$, by the rule
\begin{equation}
  \label{eq:PB}
  [f,g] = \sum_j \ptDd{f}{q_j} \ptDd{g}{p_j} -
                 \ptDd{f}{p_j} \ptDd{g}{q_j}.
\end{equation}
From this definition one may compute the \emph{fundamental} Poisson brackets
\begin{equation}
  \label{eq:fund.PB}
  [z_a, z_b] = J_{ab},
\end{equation}
where $J$ is the $2n\times2n$ matrix
\begin{equation}
  \label{eq:sympJ}
  J = \begin{pmatrix} 0 & I \\ -I & 0 \end{pmatrix}.
\end{equation}
Here $0$ and $I$ denote $n\times n$ zero and identity blocks, respectively.
The matrix $J$ is sometimes called the \emph{Poisson matrix}.

There is an \emph{existence and uniqueness theorem} to the effect that
sets of first-order ordinary differential equations have solutions,
and each solution is uniquely specified by its initial conditions%
  ~\bibnote[LM:S.1.3]{\emph{LM}, Section~1.3.}.
Hamilton's equations of motion are first order.
Now suppose a charged particle enters a beam-line element
(or collection of beam-line elements) with \emph{initial} conditions
$\V{z}^i$ and subsequently exits with \emph{final} conditions $\V{z}^f$.
Then, by the existence and uniqueness theorem, 
$\V{z}^f$ is uniquely specified by $\V{z}^i$.
Thus, there is a map $\Map{M}$, called a \emph{transfer} map,
that sends $\V{z}^i$ to $\V{z}^f$, and we write
\begin{equation}
  \label{eq:i.map.f}
  \V{z}^f = \Map{M} \V{z}^i.
\end{equation}
This relation between $\V{z}^i$ and $\V{z}^f$ is illustrated
by the picture shown in \fref{i.map.f}.

\begin{figure}[h]
  \centering
  \includegraphics*{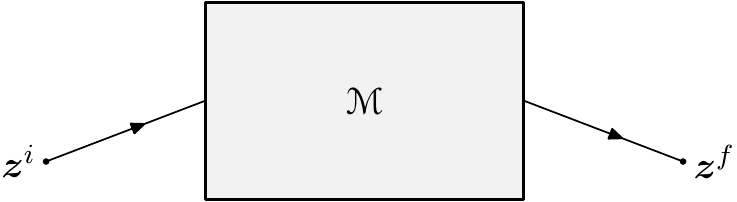}
  \caption{The transfer map \Map{M} sends initial conditions $\V{z}^i$
           to final conditions $\V{z}^f$.}
  \label{fig:i.map.f}
\end{figure}

Next suppose small changes $\ud\V{z}^i$ are made in the initial conditions.
The result will be associated small changes $\ud\V{z}^f$ in the
final conditions.  These small changes will be connected by the relations
\begin{equation}
  \label{eq:map.d}
  \ud\V{z}^f = M \ud\V{z}^i,
\end{equation}
where $M$ is the \emph{Jacobian} matrix with entries
\begin{equation}
  \label{eq:M.ab}
  M_{ab} = \ptdd{z^f_a}{z^i_b}.
\end{equation}
It can be shown that if \Map{M} is the result of integrating Hamilton's
equations of motion, then its associated Jacobian matrix $M$ will
satisfy the condition
\begin{equation}
  \label{eq:SC}
  M^T J M = J,
\end{equation}
where $M^T$ denotes the transpose of $M$.
A matrix that satisfies \eqref{SC} is said to be \emph{symplectic};
correspondingly \Map{M} is called a \emph{symplectic map}.%
  \footnote{In what follows, the letters $sp$ and $Sp$ are used as
      abbreviations for \emph{symplectic}.}
Note that in general $M$ depends on $\V{z}^i$. But $J$ does not.
Therefore \eqref{SC}, since it must hold for all $\V{z}^i$,
places strong (linear and nonlinear) restrictions on \Map{M}%
  ~\bibnote[LM:S.6.4.1]{\emph{LM}, Subsection~6.4.1.}.

Suppose the final conditions $\V{z}^f$ are to be determined by
integrating Hamilton's equations of motion \emph{numerically},
and we also wish to satisfy \eqref{SC}. A numerical integrator with
this property is called a \emph{symplectic integrator}.  In the case
of a storage ring, such as the LHC, we must integrate through thousands
of beam-line elements to integrate through even a single turn,
and we wish to integrate through a large number of turns.
Therefore, even if we wish to integrate for only a small number of turns,
we would like to be able to use
an \emph{explicit} symplectic integrator
because the numerous iterations required for
an \emph{implicit} symplectic integrator
would make the computation extremely slow.
Here we envision that one must iterate the implicit solve
to the point where convergence has been achieved to machine precision
in order to achieve symplecticity to machine precision.
Our concern is that the effort involved in iterating
implicit Runge-Kutta or implicit Gauss to machine precision
will exceed the effort required for one step of
some explicit symplectic method for $K$ or $H_R$
if such an explicit symplectic method can be found.

Explicit symplectic integrators are available if the Hamiltonian has
the form $T(p) + V(q)$.  The Hamiltonian $K$ given by \eqref{K}
is of this form if we make the approximations
\begin{equation}
  \label{eq:approx}
  \psi = 0,\ A_x = 0, \text{ and } A_y = 0,
\end{equation}
for it then takes the form
\begin{equation}
  \label{eq:K.approx}
  K = -\bigl(p_t^2 / c^2 - m^2 c^2 - p_x^2 - p_y^2\bigr)^{1/2} - q A_z.
\end{equation}
However, it can be shown that this approximation excludes
magnetic fringe-field effects, which, by the Maxwell equations,
\emph{must} occur at entry and exit of every magnetic beamline element.
It also excludes transverse electric fringe fields, which must occur
at entry and exit of all rf cavities.
To make accurate calculations that include fringe-field effects
(which can be important when nonlinear and even some linear effects
are considered), one would like to have
an \emph{explicit} symplectic integrator
that does not make the approximations in \eqref{approx}.

Remarkably, there is an explicit symplectic integrator for
the Hamiltonian $H_R$%
  ~\bibnote[LM:S.12.9]{\emph{LM}, Section~12.9.}.
But there is a caveat: One of the advertised features of symplectic
integrators is that they can be used with a rather large step size
(thereby reducing computation time) since they at least exhibit
the qualitative nature of solutions exactly.
However, there is a theorem to the effect that (for any finite step size)
symplectic integrators do not preserve the Hamiltonian, even if it has
no explicit dependence on the independent variable
($\tau$ in the case of $H_R$)%
  ~\cite{Zhong:1988:LPHJTheory}.
This may not be a serious problem in some applications of symplectic
integrators.  In the case of $H_R$, however, \eqref{H.rel.l} tells us that
failure to preserve $H_R$ means the particle mass is \emph{not} preserved.
Therefore, to preserve the particle mass to good accuracy,
which would seem highly desirable, it is necessary to employ
a sufficiently small step size, thereby making symplectic integration
in this situation relatively slow.

The rest of this chapter is devoted to exploring other possible approaches
to satisfying the symplectic condition~\eqref{SC} while at the same time
achieving improved computational speed and taking into account,
through some desired order, all linear and nonlinear effects associated
with realistic electromagnetic fields, including fringe fields and
high-order multipole fields.

%% file: LieTools.tex

\section{Lie Algebraic Concepts and Tools}
\label{sec:lie.tools}

According to \eqref{PB}, Poisson brackets obey
the \emph{antisymmetry} property
\begin{equation}
  \label{eq:PB.asymm}
  [g,f] = - [f,g].
\end{equation}
It can be verified that Poisson brackets also satisfy
the \emph{Jacobi identity}.
Let $f$, $g$, and $h$ denote any three functions on phase-space.
Then there is the identity
\begin{equation}
  \label{eq:PB.jacob}
  [f,[g,h]] + [g,[h,f]] + [h,[f,g]] = 0.
\end{equation}
As a consequence, the Poisson bracket satisfies all the requirements
for a \emph{Lie product}.
The set of all phase-space functions therefore constitutes a Lie algebra
with the Poisson bracket as the Lie product%
  ~\bibnote[LM:S.5.1]{\emph{LM}, Section~5.1.}.

Given any function $f(z,t)$, define an associated \emph{differential}
operator, denoted by $\lieop{f}$ and called a \emph{Lie operator},
by the rule
\begin{equation}
  \label{eq:lieop}
  \lieop{f} = \sum_j \ptDd{f}{q_j} \ptDd{\ }{p_j} -
                     \ptDd{f}{p_j} \ptDd{\ }{q_j}.
\end{equation}
Then, if $g(z,t)$ is any other phase-space function,
the action of $\lieop{f}$ on $g$ is defined by writing
\begin{equation}
  \label{eq:lieop.f.g}
  \lieop{f} g = \sum_j \ptDd{f}{q_j} \ptDd{g}{p_j} -
                       \ptDd{f}{p_j} \ptDd{g}{q_j} = [f,g].
\end{equation}
Thus, a Lie operator may be viewed as a Poisson bracket waiting to happen.

In general, Lie operators do not commute.
However, the commutator $\{\lieop{f},\lieop{g}\}$ of any two Lie operators
$\lieop{f}$ and $\lieop{g}$ is again a Lie operator.
Indeed, as a consequence of the Jacobi identity \eqref{PB.jacob},
the commutator $\{\lieop{f},\lieop{g}\}$ may be written in terms of
the Poisson bracket of the two underlying functions $f$ and $g$
according to the relation
\begin{equation}
  \label{eq::f.g:}
  \{\lieop{f},\lieop{g}\}
    = \lieop{f}\lieop{g} - \lieop{g}\lieop{f}
    = \lieop{[f,g]}.
\end{equation}

The relation \eqref{lieop.f.g} defines the action of $\lieop{f}$.
Powers of $\lieop{f}$ can be defined by the rules
\begin{subequations}
\label{eq:lie.pow}
\begin{align}
 & \lieop{f}^0 = { \cal{I}}\Leftrightarrow\lieop{f}^0 g = g, \\
 & \lieop{f}^1 g = [f,g], \\
 & \lieop{f}^2 g = [f,[f,g]],
\end{align}
\end{subequations}
and so on.
Now that powers of the Lie operator $\lieop{f}$ have been defined,
one may also define power series in $\lieop{f}$.  Of particular interest
is the power series associated with the exponential function by the rule
\begin{equation}
  \label{eq:lietr}
  \Lietr{f} = \lietr{f} = \sum_{m=0}^\infty \frac{1}{m!} \lieop{f}^m.
\end{equation}
This operator $\lietr{f}$, called a \emph{Lie transformation},
therefore acts on $g$ according to the relation
\begin{equation}
  \label{eq:lietr.f.g}
  \Lietr{f}g = g + [f,g] + \frac{1}{2!}[f,[f,g]] + \dotsb.
\end{equation}
In this context $\lieop{f}$ (and sometimes $f$) is called a
\emph{Lie generator}. Its importance for us lies in the fact that
\emph{any Lie transformation generates a symplectic map}.

Poisson brackets are invariant under symplectic maps%
  ~\bibnote[LM:S.6.1.2]{\emph{LM}, Section~6.1.2.}.
An important consequence of this fact is the extremely useful
similarity relation%
  ~\bibnote[LM:S.8.2]{\emph{LM}, Section~8.2.}
\begin{equation}
  \label{eq:lietr.sim}
  \Map{L}\Lietr{f}\Map[-1]{L} = \Lietr{\Map{L}f},
\end{equation}
where \Map{L} denotes any symplectic map.

Suppose the transfer map \Map{M} has the property that it maps
the origin into itself.  In other words, we assume that
$\Map{M}z$ has a Taylor expansion of the form
\begin{equation}
  \label{eq:t.map}
  z_a^f = \sum_b     R_{ab}   z_b^i
        + \sum_{bc}  T_{abc}  z_b^i z_c^i
        + \sum_{bcd} U_{abcd} z_b^i z_c^i z_d^i + \dotsb,
\end{equation}
which has no constant term.
This can always be accomplished by the use of \emph{deviation} variables.
If \Map{M} is symplectic, then $R$ must be a symplectic matrix. 
Moreover, the Taylor coefficients $T$, $U$, $\cdots$ cannot be arbitrary,
but are constrained by complicated nonlinear relations that follow
from the symplectic condition \eqref{SC}.  A \emph{truncated} Taylor expansion of a symplectic map is called a symplectic \emph{jet}. Finally, the series \eqref{t.map} \emph{cannot} in general be truncated without violating
the symplectic condition.  Therefore a symplectic jet is generally not a symplectic map.

However, there is a \emph{factorization theorem}%
  ~\bibnote[LM:S.7.6]{\emph{LM}, Section~7.6.}
to the effect that $\Map{M}$ can also be written in the Lie product form
\begin{equation}
  \label{eq:map.DF}
  \Map{M} = \Map{R} \,[\Lietr{f_3} \Lietr{f_4} \dotsb]\, \Lietr{f_1}.
\end{equation}
Here $\Map{R}$ is the linear symplectic map associated with $R$,
and the $f_m$ are \emph{homogeneous} polynomials of degree $m$.
Translations/deviations from the origin, which correspond to constant terms
added if desired to the Taylor series \eqref{t.map}, are described by $f_1$;
and the $f_3$, $f_4$ $\cdots$ describe the nonlinear terms in \eqref{t.map}.
Unlike the Taylor coefficients, there are no restrictions imposed on
the $f_m$ by the symplectic condition.

Any analytic symplectic map is uniquely specified by
a first-order polynomial $f_1$, a symplectic matrix $R$,
together with a collection of homogeneous polynomials $f_3$, $f_4$, \dots\ that
describe the nonlinear part of the map. And the converse also holds.
In addition, the (in principle infinite) product appearing
in square brackets in \eqref{map.DF} can be truncated at any stage
without violating the symplectic condition. It can be shown that
each factor in \eqref{map.DF} is a symplectic map, and the product
of any number of symplectic maps is also a symplectic map.  In what follows it is also convenient to write \eqref{map.DF} in the form
\begin{equation}
\label{eq:map.DFN}
\Map{M}=\Map{R}\Map{N}\Lietr{f_1}
\end{equation}
where $\Map{N}$, the \emph{nonlinear} part of $\Map{M}$, is given by 
\begin{equation}
\Map{N}=\Lietr{f_3} \Lietr{f_4} \cdots.
\end{equation}

%% file: SympJets.tex

\section{Symplectic Completion of Symplectic Jets}
\label{sec:symp.jets}

We know that the Lie algebra of all Lie operators, which we will call
$\ispm(2n,\mathbb{R})$, is infinite dimensional.  Correspondingly
$\ISpM(2n,\mathbb{R})$, the group of symplectic maps, is infinite
dimensional.%
  \footnote{The letters $m$ and $M$ that appear in this and the previous
    sentence are abbreviations for $map$.  The letters $i$ and $I$ are
    abbreviations for $inhomogeneous$.  By $inhomogeneous$ it is meant
    that the possibility of constant terms appearing in \eqref{t.map}
    is included.}
Indeed, the factorization \eqref{map.DFN} gives
a representation of the general analytic symplectic map.
We see that the specification of a symplectic map generally requires
an infinite number of parameters.  This fact produces an awkward situation
for human beings and computers, which can work only with a finite number
of quantities (and often only with finite precision).

An optimistic perspective on the experimental and theoretical situation,
for example in the field of accelerator physics, might be stated as follows:
\ We know that a beam transport system, accelerator, storage ring,
or any portion thereof may be described by a symplectic transfer map.
However, because we cannot measure or control electromagnetic fields exactly,
we are unsure of and unable to control exactly what this map is.
Also, since it is impossible to perform computations with an infinite number
of parameters/variables and to infinite precision, it is necessary to develop
various approximation schemes.  Thus, we are able to study computationally
(and probably theoretically) the detailed properties of only a subset
of all symplectic maps.  The hope is that if two symplectic maps are
in some sense nearly the same, then their behavior
[including, in some cases, long-term (repeated iteration) behavior]
will be in some important ways nearly the same.%
  \footnote{Note that a similar optimism is shared by practitioners of
    symplectic integration.}
Were that not true from an experimental standpoint, it would be impossible
to build satisfactory storage rings and the like.
Were that not true from a theoretical standpoint, it would be impossible
to design storage rings and the like with any assurance of satisfactory
performance.

Suppose, as an approximation, the product appearing in the square brackets
of \eqref{map.DF}, namely the map $\Map{N}$, is \emph{truncated} at
$m = \maxm$ to produce the map $\Map{N}^\trunc$ given by
\begin{equation}
  \label{eq:map.DF.tr}
  \Map{N}^\trunc
    = \Lietr{f_3}\Lietr{f_4}\dotsb\Lietr{f_\maxm}.
\end{equation}
In analogy with \eqref{map.DFN} we also make the definition
\begin{equation}
   \label{eq:map.DF.tr2}
\Map{M}^{\rm{trunc}}=\Map{R}\Map{N}^\trunc\Lietr{f_1}.
\end{equation}
The map $\Map{M}^{\rm{trunc}}$, while exactly symplectic,
requires only a finite number of parameters for its specification.
We only need store a first-order polynomial $f_1$,
a symplectic matrix $R$, and a collection of
homogeneous polynomials $f_3$, $f_4$, \dots $f_{\rm{maxm}}$.
For example, in the case of a 6-dimensional phase space,
\num{3002} parameters are required when $\maxm = 8$.

What can be said about the \emph{accuracy} of $\Map{M}^\trunc$?
Suppose the Taylor expansion \eqref{t.map} is terminated by retaining
only terms through degree $(\maxm-1)$, thereby forming
a symplectic jet that we will call $\Map{J}$.
An examination of the proof of the factorization theorem~\eqref{map.DF}
shows that a knowledge of the coefficients in $\Map{J}$
supplies just enough information to determine the ingredients
of $\Map{M}^\trunc$, and vice versa.
In other words, a knowledge of $\Map{J}$ supplies
just enough information to determine $\Map{R}$, the polynomial $f_1$,
and the polynomials $f_3$ through $f_\maxm$, and vice versa.
(In particular,
$f_3$ contributes only to $T$ terms and to terms beyond second order,
$f_4$ contributes only to $U$ terms and to terms beyond third order,
\etc)
Thus, while exactly symplectic, the map $\Map{M}^{\rm{trunc}}$
is guaranteed accurate only through terms of degree $(\maxm-1)$.
With regard to memory requirements, the storage of a jet
requires more locations because it does not exploit
the symplectic condition.
For example, \num{10296}~locations are required in the case of
6-dimensional phase space when $\maxm = 8$.

At present there are Lie-algebraic results and Truncated Power Series~(TPSA)
methods that make it possible to compute in principle the ingredients in
$\Map{M}^{\rm{trunc}}$ as given by \eqref{map.DF.tr2}, with $\maxm = 8$,
for any beam-line element or collection of beam-line elements (including
a full ring) based on field data provided numerically on a grid%
  ~\bibnote[LM:S.17ff]{\emph{LM}, Chapters~17--25.}.

What has been accomplished here?  Given a symplectic jet $\Map{J}$,
we have found a map $\Map{M}^\trunc$ that is guaranteed symplectic
and whose Taylor expansion agrees with $\Map{J}$ through terms of
degree $(\maxm-1)$. This Taylor expansion will in general contain terms
of degrees beyond $(\maxm-1)$.
It will be called the \emph{Lie symplectic completion} of $\Map{J}$.

But there is a problem:
Suppose we wish to evaluate $\Map{M}^\trunc\V{z}^i$ for some initial
condition $\V{z}^i$.  Because of the infinite series that appears in
the definition \eqref{lietr} of a Lie transformation,
the symplectic completion of $\Map{J}$ provided by $\Map{M}^\trunc$
will in general contain an \emph{infinite} number of terms.
As a consequence, the evaluation of $\Map{M}^\trunc\V{z}^i$ will
generally involve the summation of infinite series, a task that generally
lies beyond numerical methods, or is at best numerically intensive unless
the series converges rapidly.

We have seen that in principle the symplectic completion of a symplectic jet
is possible.  What we would like are other symplectic jet completions whose
actions on $\V{z}^i$ can be computed rapidly and to machine precision.
Two such methods will be described in the next two sections of this chapter.

%% file: GenFns.tex

\section{Symplectic Completion Using Generating Functions}
\label{sec:symp.gen}

We have described how in general the computation of the action of
a Lie transformation on phase space involves the summation of an
\emph{infinite} series if the symplectic condition is to be honored.
Before exploring a particular method to deal with this problem, we begin
this section by studying a simple example of what happens if only a finite
number of terms in the series expansion are employed.

As such an example consider, for a two-dimensional phase space,
the symplectic map \Map{M} given by the relation
\begin{equation}
  \label{eq:M.RN}
  \Map{M} = \Map{R}\Map{N}
\end{equation}
with linear part
\begin{equation}
  \label{eq:R.rot}
  \Map{R} = \Lietr[-\tfrac{\theta}{2}]{p^2+q^2}
\end{equation}
and nonlinear part
\begin{equation}
  \label{eq:N.e.qp2}
  \Map{N} = \Lietr{qp^2}.
\end{equation}
This map may be viewed as a toy model for the one-turn map
of a storage ring.

In this case the infinite series for \Map{R} can be summed exactly
to give the results
\begin{equation}
  \label{eq:R.rot.qp}
  Q = \Map{R}q =  q\cos\theta + p\sin\theta, \quad
  P = \Map{R}p = -q\sin\theta + p\cos\theta.
\end{equation}
And the map \Map{N} can also be evaluated exactly, since the
exponential of any monomial Lie operator can be evaluated exactly%
  ~\bibnote[LM:S.monom]{\emph{LM}, Subsections~1.2.3, 1.4.1, 1.4.2,
                        and Exercise~1.4.3.}.
For the case at hand there is the result
\begin{equation}
  \label{eq:N.qp2.qp}
  Q = \Map{N}q = q (1 - p)^2, \quad
  P = \Map{N}p = p / (1 - p).
\end{equation}
Therefore \Map{M} can also be evaluated exactly.
Note that the result \eqref{N.qp2.qp} for $P$,
and therefore the map \Map{N},
has a pole on the phase-space surface $p = 1$.

\Fref{iter.M} shows the result of applying \Map{M} repeatedly to
seven initial conditions for the case $\theta/2\pi = 0.22$.
In other words, seven initial conditions have been selected,
and their orbits have been found under the repeated action of \Map{M}.
One initial condition lies very near the origin, and its orbit
appears to lie on a closed curve that is nearly elliptical. (It would
be nearly circular had the horizontal and vertical scales been equal.)
This is to be expected because the effect of the nonlinear part \Map{N}
is so small on such orbits that they are essentially those of the
rotation map \Map{R}.  By contrast, the other initial conditions lie
successively farther from the origin, and the effect of \Map{N} becomes
ever more significant.  Their orbits appear to lie on closed curves that,
the farther they lie from the origin, the more noticeably do
nonlinearities distort them from being circular.  The origin itself
is an elliptic fixed point corresponding to a one-turn closed orbit.

\begin{figure}[t]
  \centering
  \includegraphics*{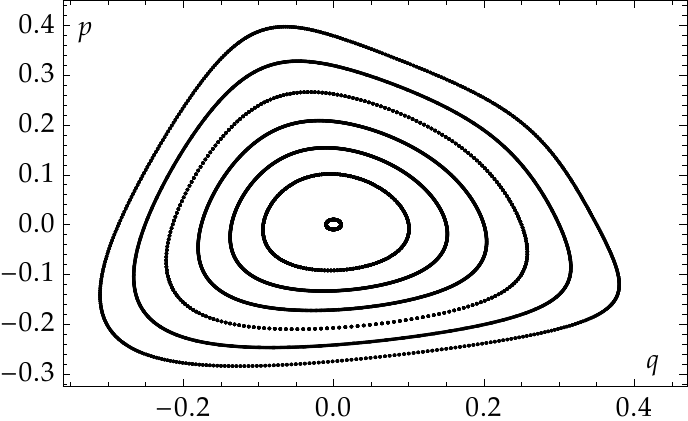}
  \caption{Phase-space portrait, in the case $\theta/2\pi = 0.22$,
    resulting from applying the map $\Map{M}$ repeatedly (2000~times)
    to the seven initial conditions
    $(q,p) = (0.01, 0)$, $(0.1, 0)$, $(0.15, 0)$, $(0.2, 0)$, $(0.25, 0)$,
    $(0.3, 0)$, and $(0.35, 0)$ to find their orbits.}
  \label{fig:iter.M}
\end{figure}

Now suppose the nonlinear map \Map{N} is \emph{truncated} to form
the map \Ntr{2} by retaining only the first \emph{two} terms
in its Taylor expansion.  In Lie form we have the result
\begin{equation}
  \label{eq:N.tr2}
  \Ntr{2} = \Map{I} + \lieop{qp^2}.
\end{equation}
This truncated map \Ntr{2} has the effect
\begin{subequations}\label{eq:N.tr2.qp}
\begin{align}
  \label{eq:N.tr2.q}
  Q &= \Ntr{2} q = (\Map{I} + \lieop{qp^2}) q
                 = q + [qp^2, q] = q - 2qp
    \Leftrightarrow Q - q = -2qp, \\
  \label{eq:N.tr2.p}
  P &= \Ntr{2} p = (\Map{I} + \lieop{qp^2}) p
                 = p + [qp^2, p] = p + p^2
    \Leftrightarrow P - p = p^2.
\end{align}
\end{subequations}
Evidently, \Ntr{2} is a symplectic jet map that
retains only terms through degree~2.
Indeed, one finds using \eqref{N.tr2.qp} the result
\begin{equation}
  \label{eq:PB.qp.2}
  [Q, P] = 1 - 4p^2 \ne 1,
\end{equation}
and therefore \Ntr{2}, while a symplectic jet map,
is as expected \emph{not} a symplectic map.%
  \footnote{It follows from \eqref{SC} that a symplectic map
            must preserve Poisson brackets, and vice versa.}


Next define a corresponding map \Mtr{2} by writing
\begin{equation}
  \label{eq:M.tr2}
  \Mtr{2} = \Map{R}\Ntr{2}.
\end{equation}
The left-hand graphic in \Fref{iter.M.tr2.psc2} shows the orbits of \Mtr{2}
for two initial conditions: one near the origin, and one quite far away.
Inspection of the figure shows that orbits are no longer distorted circles,
but instead appear to spiral into the origin.
This motion toward the origin occurs because \Ntr{2},
and consequently \Mtr{2}, is \emph{not} symplectic.

We could also retain the next term in the Taylor series for $\Lietr{qp^2}$
to form the symplectic jet map \Ntr{3},
\begin{equation}
  \label{eq:N.tr3}
  \Ntr{3} = \Map{I} + \lieop{qp^2} + \tfrac{1}{2}\lieop{qp^2}^2.
\end{equation}
It retains terms through degree~3. It is still nonsymplectic,
but more nearly symplectic than $\Ntr{2}$.
Again define a corresponding map \Mtr{3} by writing
\begin{equation}
  \label{eq:M.tr3}
  \Mtr{3} = \Map{R} \Ntr{3}.
\end{equation}
The phase-space portrait for $\Mtr{3}$ is found to be somewhat more
like that for $\Map{M}$ than that provided by $\Mtr{2}$, because
$\Mtr{3}$ is more nearly symplectic.
However, there is still substantial/disastrous nonsymplectic spiraling,
in this case out of the origin.  We see that violation of the symplectic
condition can convert the origin, initially an elliptic fixed point,
into a nonlinear attracting or a nonlinear repelling fixed point.

Suppose we know that the behavior of some dynamical system is describable
by a symplectic map.  This system may be a beam-line element, some
collection of beam-line elements, or even the one-turn map for
a storage ring.  And suppose that a truncated Taylor expansion
(symplectic jet \Map{J}) is known for this map through terms of some order.
As stated earlier, such knowledge is in fact computable using
Lie algebraic and TPSA algorithms%
  ~\bibnote[LM:S.10.5]{\emph{LM}, Section~10.5 and Chapter~39.
     Some authors refer to TPSA  as \emph{Differential Algebra} (DA).
     For an exposition of DA, see
     M.~Berz, \emph{Modern Map Methods in Particle Beam Physics},
     vol.~108 of \emph{Advances in Imaging and Electron Physics},
     Academic Press, San Diego, 1999.%
   }.
According to the factorization theorem, as far as nonlinear effects are concerned, such knowledge is equivalent to the knowledge of
a set of homogeneous polynomials $f_3$, \dots, $f_n$.
What we would like to find is a map that is symplectic, has the jet
\Map{J} through terms of degree $(n-1)$, and is relatively easy
to compute.  Because it is symplectic, its Taylor expansion must in general
also have terms beyond degree $(n-1)$.  In some way that is not yet clearly
described, we would like these additional terms to be as small as possible
while remaining consistent with the symplectic condition.
For example, their effect over the phase-space region of interest should
not be appreciably larger than the extent to which \Map{J} violates
the symplectic condition.

How can we find such symplectic maps?
We need a supply of relatively easily computed symplectic maps.
It is known that such maps can be produced
with the aid of generating functions.%
  \footnote{Symplectic completion of symplectic jets
    using a generating function was first implemented%
    ---in the context of Accelerator Physics---%
    in the Lie-algebra based accelerator design code MaryLie.}
Four types of generating functions, commonly called
$F_1(q,Q)$, $F_2(q,P)$, $F_3(p,Q)$, and $F_4(p,P)$,
are usually presented in graduate Classical Mechanics text books.
What is less familiar is that, for a $2n$ dimensional phase space,
there is in fact a $2n(4n+1)$ dimensional \emph{family} of types
of generating functions: There is a type for every $4n\times4n$
symplectic matrix%
  ~\bibnote[LM:S.ntype]{\emph{LM}, Section~6.7.}.
Among these types, we have found the so called
\emph{Poincar\'e} generating function, which we denote as $F_+$,
to be particularly attractive%
  ~\bibnote[LM:S.Poin]{\emph{LM}, Chapter~34.
     See also the article
     B.~Erd{\'{e}}lyi and M.~Berz,
     ``Optimal symplectic approximation of Hamiltonian flows,''
     \emph{Phys. Rev. Lett.}, vol.~87, 114302, Aug. 2001.
     Currently we do not find persuasive their invocation of
     the Hofer metric, but do agree (for other reasons) with their
     conclusion that use of the Poincar\'e generating function
     has several desirable features.}.

To describe the use of $F_+$, it is useful to make some additional
definitions.   In addition to the definition of $z$ given by \eqref{z.def},
which describes the collection of what we might call \emph{old} variables,
we introduce the symbol $Z$ to denote a collection of what we may call
\emph{new} variables:
\begin{equation}
  \label{eq:Z.QP}
  Z = (Z_1, Z_2, \dotsc, Z_{2n})
    = (Q_1, \dotsc, Q_n; P_1, \dotsc, P_n).
\end{equation}
We shall also need the sums and differences defined by
\begin{equation}
  \label{eq:Z.pm.z}
  \Sigma = Z + z \text{ and } \Delta = Z - z.
  \Leftrightarrow
  z = \tfrac{1}{2}(\Sigma - \Delta) \text{ and }
  Z = \tfrac{1}{2}(\Sigma + \Delta).
\end{equation}
Specifically, for future use and in the case of a two-dimensional
phase space, the relations \eqref{Z.pm.z} take the equivalent forms
\begin{subequations}
\label{eq:Z2.pm.z2}
\begin{align}
  \label{eq.Q2.pm.q2}
  \Sigma_1 &= Q + q \text{ and } \Delta_1 = Q - q
  \Leftrightarrow
  q = \tfrac{1}{2}(\Sigma_1 - \Delta_1) \text{ and }
  Q = \tfrac{1}{2}(\Sigma_1 + \Delta_1), \\
  \label{eq.P2.pm.p2}
  \Sigma_2 &= P + p \text{ and } \Delta_2 = P - p
  \Leftrightarrow
  p = \tfrac{1}{2}(\Sigma_2 - \Delta_2) \text{ and }
  P = \tfrac{1}{2}(\Sigma_2 + \Delta_2).
\end{align}
\end{subequations}
Finally, we will need a collection of $2n$ \emph{auxiliary} variables,
which we will call $u$:
\begin{equation}
  \label{eq:u.def}
  u = (u_1, u_2, \dotsc, u_{2n}).
\end{equation}

Now let $F_+(u,t)$ be any function of $u$ and perhaps the
independent/time  variable $t$.
We define its action on phase space by the rule
\begin{equation}
  \label{eq:D.def}
  \Delta = J \partial_u F_+ |_{u=\Sigma}.
\end{equation}
Note that this definition specifies a relation
between $\Delta$ and $\Sigma$,
which in turn, when the relations on the right-hand side of \eqref{Z.pm.z}
are taken into account, specifies a relation between $Z$ and $z$.
By the general theory of generating function machinery, this relation
between $Z$ and $z$ is guaranteed to be a symplectic map
for \emph{any} choice of $F_+(u,t)$.

To see how this mathematical machinery works in some detail, let us
apply it to a simple example in two-dimensional phase-space.
Suppose $F_+$ is the cubic monomial
\begin{equation}
  \label{eq:F+3}
  F_+(u) = F^3_+(u) = -\tfrac{1}{4} u_1 {u_2}^2.
\end{equation}
We then compute
\begin{equation}
  \label{eq:JdF}
  J\partial_u F^3_+
    = \begin{pmatrix} 0 & 1 \\ -1 & 0 \end{pmatrix}
      \begin{pmatrix} -\tfrac{1}{4}{u_2}^2 \\[0.7ex]
                      -\tfrac{1}{2}u_1u_2 \end{pmatrix}
    = \begin{pmatrix} -\tfrac{1}{2}u_1u_2 \\[0.7ex]
                      \,\tfrac{1}{4}{u_2}^2 \end{pmatrix},
\end{equation}
and hence
\begin{equation}
  \label{eq:JdF.u}
  J\partial_u F^3_+|_{u=\Sigma}
  {\;} = \begin{pmatrix} -\tfrac{1}{2}\Sigma_1\Sigma_2 \\[0.7ex]
                          \tfrac{1}{4}{\Sigma_2}^2 \end{pmatrix}.
\end{equation}
From \eqref{D.def} and \eqref{JdF.u} it follows that
\begin{equation}
  \label{eq:D.12}
  \Delta_1 = -\tfrac{1}{2}\Sigma_1\Sigma_2,
  \text{ and }
  \Delta_2 =  \tfrac{1}{4}{\Sigma_2}^2.
\end{equation}
Finally, employ the relations \eqref{Z2.pm.z2} in \eqref{D.12} to obtain
the relations
\begin{subequations}
\label{eq:QP.qp}
\begin{align}
  \label{eq:Q.q}
  Q - q = -\tfrac{1}{2} (Q + q) (P + p)
    & \Leftrightarrow Q = q - \tfrac{1}{2} (Q + q) (P + p), \\[0.5ex]
  \label{eq:P.p}
  P - p = \tfrac{1}{4}(P+p)^2
    & \Leftrightarrow P = p + \tfrac{1}{4} (P + p)^2.
\end{align}
\end{subequations}
The two equations on the right-hand sides of \eqref{QP.qp} specify
a relation between $Z$ and $z$.

As is the case with the use of any generating function, the above relation
between $Z$ and $z$ is \emph{implicit}.
We can \emph{begin} to make it \emph{explicit} by seeking a Taylor expansion
using iteration:  For the zeroth iteration, make the Ansatz
\begin{subequations}
\label{eq:QP.iter}
\begin{align}
  \label{eq:QP.ansatz}
  Q = q\text{ and }
  P = p. \\
\intertext{%
Now substitute this Ansatz into the right-hand sides of \eqref{QP.qp}
to yield for the first iteration the results}
  \label{eq:QP.a1}
  Q = q - 2 q p\text{ and }
  P = p + p^2.
\end{align}
\end{subequations}
Observe that the results \eqref{QP.a1} agree
with the jet results \eqref{N.tr2.qp}.
We have found a symplectic map whose jet through terms of second order
agrees with \Ntr{2}.

\begin{figure}[t]
  \centering
  \includegraphics*{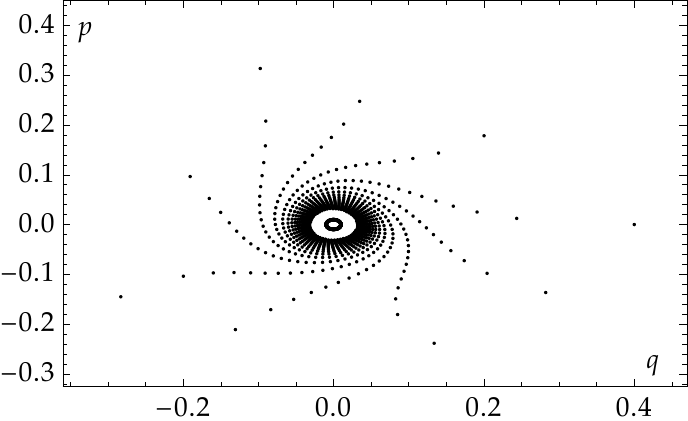}
  \hfill
  \includegraphics*{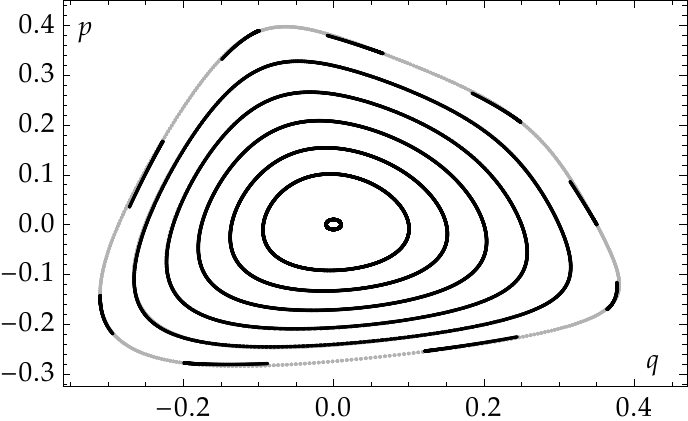}
  \caption{Phase-space portraits, in the case $\theta/2\pi = 0.22$.
    Left:~The resulting of applying the map \Mtr{2} repeatedly
    (\num{1000}~times) to the two initial conditions $(q, p) = (0.01, 0)$
    and $(0.4, 0)$. These orbits spiral into the origin.
    Right:~The result of applying the map \Mpsc{2} repeatedly
    (\num{2000}~times) to the seven initial conditions
    $(q, p) = (0.01, 0)$, $(0.1, 0)$, $(0.15, 0)$, $(0.2, 0)$,
             $(0.25, 0)$, $(0.3, 0)$, and $(0.35, 0)$.
    Light gray curves in the background indicate the exact result.}
  \label{fig:iter.M.tr2.psc2}
\end{figure}

It happens that for this example the implicit equations on the right-hand
sides of \eqref{QP.qp} can be made explicit by algebraic manipulation.
The equation on the right-hand side of \eqref{P.p} is quadratic in $P$,
and on choosing the negative square root
(the solution for which $p$ vanishing implies that $P$ also vanishes),
we obtain the relation
\begin{subequations}
\label{eq:N.psc.qp}
\begin{equation}
  \label{eq:N.psc.p}
  P = - (p - 2) - 2 \sqrt{1 - 2p}.
\end{equation}
Once $P$ is known, the equation on the right-hand side of \eqref{Q.q}
is linear in $Q$ and has the immediate solution
\begin{equation}
  \label{eq:N.psc.q}
  Q = \frac{q\sqrt{1 - 2p}}{2 - \sqrt{1 - 2p}}.
\end{equation}
\end{subequations}
%
%
We see that, in this case, the use of $F_+$ produces a map that has
a square-root branch point on the phase-space surface $p=1/2$ and
a pole at $p=-3/2$.  Note that the branch-point singularity lies closer
to the origin than does the pole of the exact map, which we have noted
lies on the surface $p=1$.

Let \Npsc{2} be the map given by \eqref{N.psc.qp}.
One can verify by direct computation that it has the property
\begin{equation}
  [Q, P] = 1,
\end{equation}
and hence the map \Npsc{2} is exactly symplectic, as desired and expected.
The reader can also verify that the terms through degree~$2$ in the Taylor
expansions of \eqref{N.psc.qp} agree with the terms in \eqref{N.tr2.qp}.
This is just the result \eqref{QP.a1} that we have already found by
iteration.%
  \footnote{Surprisingly, the terms through degree~$3$ in the Taylor
            expansions of \eqref{N.psc.qp} \emph{agree} with the terms
            generated by applying \eqref{N.tr3} to $(q,p)$.
            This happens due to the second relation in \eqref{F_+.34}
            and our tacit assumption that $f_4=0$.}
We may therefore say that \Npsc{2} is the
\emph{Poincar\'e symplectic completion}
of the degree-two symplectic jet map \Ntr{2} given by \eqref{N.tr2.qp}.
Correspondingly, suppose we define the associated map \Mpsc{2} by the
relation
\begin{equation}
  \label{eq:M.psc2}
  \Mpsc{2} = \Map{R}\Npsc{2}.
\end{equation}
We expect it to be symplectic because it is the product
of two symplectic maps.

The right-hand graphic in \Fref{iter.M.tr2.psc2} shows the result
of applying \Mpsc{2} repeatedly to seven initial conditions for
the case $\theta/2\pi = 0.22$.
Note that orbits generated by \Mpsc{2} exhibit no spurious spiraling
towards or away from the origin.
Moreover, comparison with the background light gray curves showing
the exact result reveals that the orbits closely agree in shape,
but there is some difference in phase advance.
It seems remarkable that the relatively meager information about
$\Map{N}$ present in \Ntr{2} suffices, after
Poincar\'e symplectic completion has been performed,
to give such good agreement.


The reader may wonder how we knew to make the inspired choice \eqref{F+3}
for $F_+^3(u)$.  Suppose we make for $F_+(u)$ the expansion
\begin{equation}
  \label{eq:F+}
  F_+(u) = \sum_{m=3}^\maxm F_+^m(u),
\end{equation}
where the $F_+^m(u)$ are homogeneous polynomials of degree $m$.
Then there are formulas that determine the $F_+^m(u)$ in terms of
the $f_m(u)$.  For example, there are the relations
\begin{equation}
  \label{eq:F_+.34}
  F_+^3(u) = -\tfrac{1}{4} f_3(u) \text{ and }
  F_+^4(u) = -\tfrac{1}{8} f_4(u).
\end{equation}
Observe that if we assume $F_+(u)$ has \emph{only} a $F_+^3(u)$ component,
then the associated map will have an $f_3$ given by \eqref{F_+.34}
and a \emph{vanishing} $f_4$.
(This is one of the virtues of the Poincar\'e generating function.)
In general the $f_{>4}$ will not vanish.
Results for the $F_+^m(u)$ in terms of the $f_m(u)$ are known through
order $m = 8$, but become increasingly complicated as $m$ increases
beyond $m = 4$%
  ~\bibnote[LM:S.34.4.]{\emph{LM}, Section 34.4.}.
However, it is also possible to proceed \emph{without} these formulas,
thereby bypassing their complications.
We will next illustrate how to do so for our simple example.

Suppose only the functions $f_3$, $f_4$, \dots, $f_\maxm$ are known.
For our example we know from \eqref{N.e.qp2} that $f_3 = qp^2$
and $(\maxm-1) = 2$.
Next compute the set of Taylor series through terms of degree $(\maxm-1)$
for the jet \Ntr{(\maxm-1)}.
For our example this set is given by \eqref{N.tr2.qp}.
In this set, replace $Q,q$ and $P,p$ by their representations in terms of
$\Sigma$ and $\Delta$ using, for this example, the relations given on
the right-hand sides of \eqref{Z2.pm.z2}.  So doing, for the relations
appearing on the far-right sides of \eqref{N.tr2.qp}, yields the results
\begin{subequations}
\label{eq:N.tr2.SD}
\label{eq:N.tr2.D12}
\begin{align}
  \label{eq:N.tr2.D1}
  \Delta_1 &= -2\,\bigl[\tfrac{1}{2}(\Sigma_1 - \Delta_1)\,
                        \tfrac{1}{2}(\Sigma_2 - \Delta_2)\bigr]
            = -\tfrac{1}{2}(\Sigma_1 - \Delta_1)
                           (\Sigma_2 - \Delta_2), \\[0.5ex]
  \label{eq:N.tr2.D2}
  \Delta_2 &= \bigl[\tfrac{1}{2}(\Sigma_2 - \Delta_2)\bigr]^2
            = \tfrac{1}{4}(\Sigma_2 - \Delta_2)^2.
\end{align}
\end{subequations}

Approximately solve these equations for the quantities
$\Delta$ in terms of the quantities $\Sigma$.
Do so in the form of a Taylor series in $\Sigma$ truncated beyond
terms of degree $(\maxm-1)$, which can be done by iteration:
For the zeroth iteration make the Ansatz
\begin{subequations}
\begin{align}
  \label{eq:D12.ansatz}
  \Delta_1 = 0 &\text{ and }
  \Delta_2 = 0.
 \\\intertext{%
Now substitute this Ansatz into the right-hand sides of \eqref{N.tr2.D12}
to obtain, through terms of degree~$2$, the Taylor expansion}
  \label{eq:D12.a1}
  \Delta_1 = -\tfrac{1}{2}\Sigma_1\Sigma_2 &\text{ and }
  \Delta_2 =  \tfrac{1}{4}{\Sigma_2}^2.
\end{align}
\end{subequations}
In this case the iteration process is finished because
$(\maxm - 1) = 3 - 1 = 2.$
Observe that the relations \eqref{D12.a1} agree with the relations
\eqref{D.12}! The transition from \eqref{D.12} to the right-hand sides
of \eqref{QP.qp} now proceeds as before.
We have found, \emph{directly} from the jet \Ntr{(maxm-1)},
the relations that would have flowed from the use of the related $F_+$.
And we have therefore identified (in implicit form) the desired
symplectic map.
Note also that the operations we have just performed involve only
well-defined polynomial manipulations, and therefore can be performed
on a computer using TPSA routines.

There remains the problem of converting the implicit results given
by \eqref{D.def} to explicit results for $Z$ in terms of $z$. For our
simple example we were able to do so by solving a quadratic equation.
For most applications, however, the equations to be solved are much more
complicated, and must be handled numerically.
That is, given $z$ as a collection of numbers, we would like to find
the associated collection of numbers $Z$.
This can be done by simple iteration or by use of Newton's method.
In either case the process can be started using jet results.

As an example of the use of simple iteration, the equations on the right-hand
sides of \eqref{QP.qp} may be converted into the iteration rule
\begin{subequations}
\label{eq:QnPn.iter}
\begin{align}
\label{eq:Qn.iter}
  Q^{[n+1]} &= q - \tfrac{1}{2}(Q^{[n]} + q)(P^{[n]}+p), \\[0.5ex]
\label{eq:Pn.iter}
  P^{[n+1]} &= p + \tfrac{1}{4}(P^{[n]} + p)^2.
\end{align}
\end{subequations}
Using the jet results \eqref{N.tr2.qp}, one may begin the iteration
optimally with the values
\begin{equation}
  \label{eq:QP.0}
  Q^{[0]} = q - 2qp \text{ and }
  P^{[0]} = p +p^2.
\end{equation}
Observe that in general the quantities to be evaluated numerically
at each step are polynomials, and therefore this evaluation is quite fast.

For example, consider the case
\begin{equation}
  \label{eq:qp.0}
  q = -0.3 \text{ and }
  p = -0.2,
\end{equation}
which is a point near the boundary of \fref{iter.M}.
Then use of \eqref{N.psc.qp} shows that in this case we hope
to find the results
\begin{equation}
  \label{eq:qp.inf}
  Q^\infty = \num{-0.43458829768152063}\cdots \text{ and }
  P^\infty = \num{-0.16643191323984635}\cdots.
\end{equation}

\begin{table}[t]
  \centering
  \caption{Convergence of $Q^{[n]},P^{[n]}$ and 
           $\Delta Q^{[n]},\Delta P^{[n]}$ as a function of $n$
           using simple iteration.}
  \label{tbl:QP.iter}
  \sisetup{table-number-alignment = left,
           table-figures-integer = 2
  }\small
  \begin{tabular}{
    S[table-figures-decimal =  1]
    S[table-figures-decimal = 16]
    S[table-figures-decimal = 16]
    S[
      table-figures-decimal  = 2,
      table-figures-exponent = 2,
      round-mode = figures,
      round-precision = 3
    ]
    S[
      table-figures-decimal  = 2,
      table-figures-exponent = 2,
      round-mode = figures,
      round-precision = 3
    ]
  }
  \toprule
  \multicolumn{1}{c}{$n$}   &
  \multicolumn{1}{c}{$Q^{[n]}$} &
  \multicolumn{1}{c}{$P^{[n]}$} &
  \multicolumn{1}{c}{$\Delta Q^{[n]}$} &
  \multicolumn{1}{c}{$\Delta P^{[n]}$} \\
  \midrule
   0 & -0.4200000000000000 & -0.1600000000000000 & -1.459e-2  & -6.432e-3  \\
   1 & -0.4295999999999999 & -0.1676000000000000 & -4.988e-3  &  1.168e-3  \\
   2 & -0.4341004800000000 & -0.1662175600000000 & -4.878e-4  & -2.144e-4  \\
   3 & -0.4344202432902144 & -0.1664711746869116 & -1.681e-4  &  3.926e-5  \\[-0.6ex]
 \vdots & \vdots           & \vdots              & \vdots     & \vdots     \\[0.3ex]
  17 & -0.4345882976815126 & -0.1664319132398483 & -7.994e-15 &  1.943e-15 \\
  18 & -0.4345882976815200 & -0.1664319132398461 & -6.661e-16 & -2.776e-16 \\
  19 & -0.4345882976815204 & -0.1664319132398465 & -1.665e-16 &  1.388e-16 \\
  20 & -0.4345882976815207 & -0.1664319132398464 &  5.551e-17 &  5.551e-17 \\
  \bottomrule
\end{tabular}
\end{table}
\begin{table}[t]
  \centering
  \caption{Convergence of $Q^{[n]},P^{[n]}$ and
           $\Delta Q^{[n]},\Delta P^{[n]}$ as a function of $n$ using
           Newton's method.}
  \label{tbl:NQP.iter}
  \sisetup{table-number-alignment = left,
           table-figures-integer = 2
  }\small
  \begin{tabular}{
    S[table-figures-decimal =  1]
    S[table-figures-decimal = 16]
    S[table-figures-decimal = 16]
    S[
      table-figures-decimal  = 2,
      table-figures-exponent = 2,
      round-mode = figures,
      round-precision = 3
    ]
    S[
      table-figures-decimal  = 2,
      table-figures-exponent = 2,
      round-mode = figures,
      round-precision = 3
    ]
  }
  \toprule
  \multicolumn{1}{c}{$n$}   &
  \multicolumn{1}{c}{$Q^{[n]}$} &
  \multicolumn{1}{c}{$P^{[n]}$} &
  \multicolumn{1}{c}{$\Delta Q^{[n]}$} &
  \multicolumn{1}{c}{$\Delta P^{[n]}$} \\
  \midrule
   0 & -0.4200000000000000 & -0.1600000000000000 & -1.459e-2  & -6.432e-3 \\
   1 & -0.4345349317899958 & -0.1664406779661017 & -5.337e-5  &  8.765e-6 \\
   2 & -0.4345882979751494 & -0.1664319132560776 &  2.936e-10 &  1.623e-11 \\
   3 & -0.4345882976815208 & -0.1664319132398464 &  1.110e-16 &  8.327e-17 \\
   4 & -0.4345882976815208 & -0.1664319132398464 &  1.110e-16 &  5.551e-17 \\
  \bottomrule
  \end{tabular}
\end{table}

For the case \eqref{qp.0}, \tref{QP.iter} displays simple iteration results
near the beginning and end of the iteration process.  We see that the point
$Q^\infty,P^\infty$ is, as desired, an \emph{attractor} for the simple
iteration process. Convergence to machine precision has been achieved
after about 20~iterations.
However, the simple iteration process converges rather slowly.%
  \footnote{The iteration process does converge more rapidly for
      $q,p$ values that lie closer to the origin, but the improvement
      is not impressive.  For example, after reducing the distance from
      the origin by a factor of $5$, at least $12$ iterations are still
      required to achieve convergence to machine precision.}
Let us examine the convergence rate.  Define errors
$\Delta Q^{[n]},\Delta P^{[n]}$ by the rules
$\Delta Q^{[n]} = Q^{[n]} - Q^\infty$ and
$\Delta P^{[n]} = P^{[n]} - P^\infty$.
Examination of the error columns in \tref{QP.iter} shows that there are
the results
$|\Delta Q^{[n+1]} / \Delta Q^{[n]}| \sim 0.2$ and
$|\Delta P^{[n+1]} / \Delta P^{[n]}| \sim 0.2$,
and therefore the errors decrease \emph{geometrically} with each iteration
by factors that are not significantly less than $1$.
The number of correct digits increases roughly \emph{linearly}
with the number of iterations

We would like an iteration process that is more rapidly convergent.
Solution of the equations on the right-hand sides of \eqref{QP.qp}
can be converted into a fixed-point problem, and this problem can
be solved by Newton's method.  The results of so doing for the problem
at hand are shown in \tref{NQP.iter}.  Evidently with Newton's method
convergence to machine precision has been achieved with $3$ iterations.
And for the errors we find the results
$|\Delta Q^{[n+1]} / (\Delta Q^{[n]})^2| \sim 0.2$ and
$|\Delta P^{[n+1]} / (\Delta P^{[n]})^2| \sim 0.1$.
Therefore the convergence is \emph{quadratic} as expected for
Newton's method.  The number of correct digits roughly \emph{doubles}
from one iteration to the next.

%% file: Cremona.tex

\section{Symplectic Completion Using Cremona Maps}
\label{sec:cremona}

The previous section illustrated how one may begin with a symplectic jet
$\cal J$ truncated at terms beyond degree $(\maxm-1)$ and then,
by use of a suitable Poincar\'e generating function, add
to $\cal J$ terms of degree $\maxm$ and higher to produce an exactly
symplectic map.
In effect, that approach usually adds an \emph{infinite} number of terms
beyond degree $(\maxm-1)$, because the resulting map generally contains
singularities.  However, as seen for phase-space regions of physical
interest, these additional terms seem to have little (no deleterious)
effect beyond achieving symplectification.

In this section we describe how one may achieve symplectification
by adding only a \emph{finite} number of terms beyond degree $(\maxm-1)$.
The result is a map that is both \emph{polynomial} and exactly symplectic.
We call such maps \emph{Cremona} maps.
For simplicity, as was done in the previous section, we restrict our
discussion to the case of one degree of freedom%
  ~\bibnote[TMCA:init]{%
     Detailed background material for this section is most easily found
     on the Web:~%
     D.\,T. Abell, \emph{Analytic Properties and Cremona Approximation
     of Transfer Maps for Hamiltonian Systems}, PhD dissertation,
     University of Maryland, College Park, 1995.
     \urlprefix\url{%
       https://www.radiasoft.net/wp-content/uploads/2021/01/thesis_2e.pdf}.
     In subsequent citations it will be referred to as \emph{TMCA}.
     See also the chapter
     ``Symplectic maps and computation of orbits
           in particle  accelerators'' by A.\,J. Dragt and D.\,T. Abell,
     in the book \emph{Integration Algorithms and Classical Mechanics}
     (J.\,E. Marsden, G.\,W. Patrick, and W.\,F. Shadwick, eds.),
     vol.~10 of \emph{Fields Inst. Comm.}, (Providence, Rhode Island),
     pp.~59--85, American Mathematical Society, 1996.
     See further
     S.~Blanes,
     ``Symplectic maps for approximating polynomial Hamiltonian systems,''
     \emph{Phys.\ Rev.\ E}, vol.~65, 056703, May 2002.
     For a history of the term \emph{Cremona maps}, and for a fuller
     discussion of the concepts of \emph{kicks} and \emph{jolts}
     than we present here (including the extension to two and three
     degrees of freedom), see \emph{TMCA}, Chapters~10 and 11.%
}.

As Cremona maps are symplectic, one can use them to approximate the behavior
of Hamiltonian systems; as they are polynomial, one can compute them
rapidly and exactly. But how might one construct such maps?
To answer this question, we consider first the Lie transformation of a
polynomial function of $q$ alone; in other words, a map of the form
$\Lietr{g(q)}$ with $g$ an arbitrary polynomial in $q$.
Using \eqref{lietr.f.g} and \eqref{PB}, one may compute the action
of such a map to be
\begin{equation}
  \label{eq:kick}
  \Lietr{g(q)} \begin{pmatrix} q \\ p \end{pmatrix}
  = \begin{pmatrix} q \\ p + \ptdd{g}{q} \end{pmatrix},
\end{equation}
which is necessarily a polynomial symplectic map.
Because this map changes only the momentum, we call it a \emph{kick map}
and refer to the corresponding Lie generator $g$ as a \emph{kick}.
Now suppose we choose any linear symplectic map \Map{L}
and use it to form a more general map $\Lietr{\Map{L}g(q)}$.
With the aid of \eqref{lietr.sim} one may write
\begin{equation}
  \label{eq:jolt}
  \Lietr{\Map{L}g(q)}=\Map{L}\Lietr{g(q)}\Map[-1]{L}.
\end{equation}
Then the linearity of \Map{L}
assures us that maps of the form \eqref{jolt} are also necessarily
polynomial symplectic maps.
We call such a generalized kick map a \emph{jolt map}
and its Lie generator, $\Map{L}g$, a \emph{jolt}.
What we have learned here is that jolt maps---defined by an \Map{L}
and a $g(q)$---can supply us with an endless stream of Cremona maps.

Later in this paper, we shall make use of jolt maps to construct
a Cremona symplectification for our example map~\eqref{M.RN}.
To do so, we first need to develop some concepts and tools.

In the vector space of dynamical polynomials
(\ie~all polynomials on phase space),
we define a set of general basis monomials of degree $l$ by the rule
\begin{equation}
  \label{eq:Gdef}
  \Gl_r(z) = \frac{q^{l-r} p^{r}}
                  {\sqrt{(l-r)! r!}}.
\end{equation}
And for the basis monomial in $q$ alone, we write $\Ql = \Gl_0$; thus,
\begin{equation}
  \label{eq:Qdef}
  \Ql(q) = \frac{q^{l}}{\sqrt{l!}}.
\end{equation}
In addition, we introduce a (very special) inner product $\braket{\,,\,}$
defined by the rule
\begin{subequations}\label{eq:isp}
\begin{equation}
  \label{eq:isp1}
  \braket{\Gl_r, G^{(l')}_{r'}} = \delta_{ll'} \delta_{rr'}.
\end{equation}
We see that with respect to this inner product, the $\Gl_r$ constitute
an orthonormal basis for the space of dynamical polynomials. Now suppose
we have dynamical polynomials $f=\sum_{lr} f_{lr} \Gl_r$ and
$g=\sum_{lr} g_{lr} \Gl_r$. We extend the inner product \eqref{isp1},
to the entire vector space of dynamical polynomials by defining
\begin{equation}
  \label{eq:isp2}
  \braket{f,g} = \sum_{lr} f_{lr}\, g_{lr}.
\end{equation}
\end{subequations}

For our purposes, the essential feature of what we call the
\emph{invariant scalar product}, \eqref{isp}, is that
any transformation belonging to the \U{1} subgroup of \SpR{2}
leaves this inner product unchanged%
  ~\bibnote[LM:S.7.3]{\emph{LM}, section~7.3.}.
That subgroup is the group of plane rotations, and hence
\[
  \braket{\Map{L}f, \Map{L}g} = \braket{f,g}
\]
for any \Map{L} having the form, \eqref{R.rot}, of a plane rotation.
Because the jolt $\Map{L}g(q)$ is generally a polynomial in both $q$
and $p$ (recall \eqref{R.rot.qp}), we may hope that a set of jolts 
$\{\Map{L}_j\Ql(q)\}$ can span the space of relevant dynamical polynomials.

We have two other concepts to introduce. The first is that of
\emph{sensitivity vectors} $\sigma^r$, which have components%
  \footnote{The $\sigma^r_j$ do depend on $l$, but we have suppressed
            this index to avoid notational clutter.}
\begin{equation}
  \label{eq:sigma.r}
  \sigma^r_j = \braket{\Gl_r, \Map{L}_j \Ql}.
\end{equation}
These components measure the content of each jolt $\Map{L}_j \Ql$
within the given monomial $\Gl_r$.
The second concept is that of the \emph{Gram matrix} $\Graml$,
which has components
\begin{equation}
  \label{eq:gram.rs}
  \Graml_{rs} = \frac{1}{N} \sum_{j=1}^N \sigma^r_j\, \sigma^s_j
            \equiv \{\sigma^r, \sigma^s\},
\end{equation}
where we have introduced a weighted scalar product denoted $\set{\,,\,}$.%
  \footnote{In the case of two or three degrees of freedom,
            this becomes $\sum_j w_j \sigma^r_j \sigma^s_j$,
            with weights $w_j$ differing from $1/N$.}
This symmetric matrix, which measures the uniqueness, or linear independence,
of the different sensitivity vectors, depends only on one's choice of
$\Map{L}_j$.

With the above concepts and tools in hand, let us return to the example
nonlinear Lie generator $qp^2$ of \eqref{N.e.qp2}. We ask ourselves,
``How can one express this generator as a linear combination of jolts?''
Or the slightly more general question: How do we determine
$N$ jolts $\Map{L}_j \Ql[3]$
together with associated \emph{jolt strengths} $a^{(3)}_j$
so as to obtain a \emph{jolt decomposition},
\begin{equation}
  \label{eq:jolt.decomp}
  f_3(q,p) = \frac{1}{N} \sum_{j=1}^N a^{(3)}_j \Map{L}_j \Ql[3],
\end{equation}
for any homogeneous dynamical polynomial $f_3$ of degree three?
Because the monomials $\Gl[3]_r$ form a basis for such polynomials,
we may, with the use of \eqref{isp1}, write
\begin{equation}
  \label{eq:monom.decomp}
  f_3 = \sum_{r=0}^3 c^{(3)}_r \Gl[3]_r,
  \text{ where }
  c^{(3)}_r = \braket{\Gl[3]_r,f_3}.
\end{equation}
Now insert the jolt decomposition \eqref{jolt.decomp} into
the latter equality. We find that
\begin{equation}
  \label{eq:c3r.1}
  c^{(3)}_r = \frac{1}{N}
              \sum_{j=1}^N a^{(3)}_j \braket{\Gl[3]_r, \Map{L}_j \Ql[3]}
            = \frac{1}{N}
              \sum_{j=1}^N a^{(3)}_j \sigma^r_j,
            = \{a^{(3)}, \sigma^r\}.
\end{equation}
Since no component of $a^{(3)}$ that lies orthogonal to $\sigma^r$
can contribute to $c^{(3)}_r$, we make the \emph{Ansatz} that
the vector of jolt strengths, $a^{(3)}$, must be a linear combination
of the sensitivity vectors;
thus $a^{(3)} = \sum_s \alpha^{(3)}_s \sigma^s$.
Inserting this expansion into \eqref{c3r.1}, we obtain
\begin{equation}
  \label{eq:c3r.2}
  c^{(3)}_r = \sum_{s=0}^3 \alpha^{(3)}_s \{\sigma^s, \sigma^r\}
            = \sum_{s=0}^3 \Graml[3]_{rs}\,\alpha^{(3)}_s;
  \text{ hence }
  c^{(3)} = \Graml[3]\,\alpha^{(3)}.
\end{equation}
This result tells us that in order to compute $\alpha^{(3)}$---and
hence the jolt strengths $a^{(3)}$---we require a non-singular
Gram matrix $\Graml[3]$.
We may then compute $\alpha^{(3)} = \Graml[3]^{-1} c^{(3)}$.

One may parameterize the $\Map{L}_j$ by rotation angles $\theta_j$.
It is then possible to compute the sensitivity vectors analytically,
with result
\begin{equation}
  \label{eq:sigma.rj}
  \sigma^r_j = \binom{3}{r}^{1/2} c_j^{3-r} s_j^r,
\end{equation}
where $c_j = \cos\theta_j$ and $s_j = \sin\theta_j$.
One then obtains the Gram matrix elements in the explicit form%
  ~\bibnote[TMCA:S.16.1.1]{\emph{TMCA}, Section~16.1.1}
\begin{equation}
  \label{eq:gram.rs.d1}
  \Graml[3]_{rs} = {\biggl[\binom{3}{r}\binom{3}{s}\biggr]}^{1/2}
                   \frac{1}{N}\sum_j c_j^{6-(r+s)} s_j^{r+s}.
\end{equation}
The questions that remain are
(i)~what is the best choice of angles $\theta_j$,
and (ii)~how many do we need?
It seems reasonable to be democratic about our choice of angles,
and analysis has indeed shown evenly-spaced angles to be optimal%
  ~\bibnote[TMCA:S.16.1.3]{\emph{TMCA}, Section~16.1.3}.
A simple dimension-counting argument tells us that for the case $l=3$,
the number of jolts cannot be less than \num4.
However, an analysis of \eqref{gram.rs.d1}%
  ~\bibnote[TMCA:S.16.1.4ff]{\emph{TMCA}, Sections~16.1.4 and 16.1.6},
or direct numerical computation, shows that
four evenly-spaced angles yield a singular $\Graml[3]$;
but use of five evenly-spaced angles does not.

Let us review our progress so far: We have the third-order Lie generator
$qp^2$, which we want to decompose into a linear combination of jolts,
as in \eqref{jolt.decomp}. Using the evenly-spaced angles
$\theta_j = 2\pi{}j/5$, we construct, \cf~\eqref{R.rot},
the five rotation maps
\(
  \Map{L}_j = \Map{R}(\theta_j),
\)
with $j\in\set{0,\dotsc,4}$.
We also construct the Gram matrix \eqref{gram.rs.d1},
sensitivity vectors \eqref{sigma.rj},
and the jolt strengths $a^{(3)} = \sum_r \alpha^{(3)} \sigma^r$,
where $\alpha^{(3)} = \Graml[3]^{-1} c^{(3)}$.
This allows us, finally, to construct the jolt decomposition
\[
  qp^2 \approx
  - 0.6532\,\Map{L}_0\Ql[3] + 0.8936\,\Map{L}_1\Ql[3]
  - 0.5670\,\Map{L}_2\Ql[3] - 0.5670\,\Map{L}_3\Ql[3]
  + 0.8936\,\Map{L}_4\Ql[3].
\]
The last step is to split this decomposition into five individual jolt maps.
We thereby achieve an approximation to \Map{N}
in the form of a Cremona map,
\begin{multline}
  \label{eq:N.cr2}
  \Ncr =
  \Map{L}_0\lietr{-0.6532\Ql[3]}\Map{L}_0^{-1}\,
  \Map{L}_1\lietr{ 0.8936\Ql[3]}\Map{L}_1^{-1}\,
  \Map{L}_2\lietr{-0.5670\Ql[3]}\Map{L}_2^{-1}\\
  \times
  \Map{L}_3\lietr{-0.5670\Ql[3]}\Map{L}_3^{-1}\,
  \Map{L}_4\lietr{ 0.8936\Ql[3]}\Map{L}_4^{-1}.
\end{multline}
We make two observation about this form. First, the error made
by this approximation is,
according to the Baker-Campbell-Hausdorff theorem%
  ~\bibnote[LM:S.3.7.3]{\emph{LM}, Section~3.7.3},
determined by commutators and multiple commutators of the different
jolts, which in this case have degree~\num4 and higher,
\ie, starting at one degree higher than the jolts themselves.
The hope is that those higher-degree terms effectively added to \Map{N}
by our Cremona factorization do not damage the dynamics.
Second, the order of the factors in \eqref{N.cr2} is not prescribed.
We may reorder them without changing the degree of approximation.

\Fref{iter.C.5j} shows the result of applying $\Mcr = \Map{R}\Ncr$
repeatedly to seven initial conditions for the case $\theta/2\pi = 0.22$.
The two graphics in that figure correspond to different orderings of
the factors in \eqref{N.cr2}.
To facilitate comparison, light gray curves in the background indicate
the corresponding exact results.
Evidently, the order of factors can have a dramatic impact on
the result's absolute accuracy.
In the left-hand graphic, the original outer curve is replaced
by five islands, and the next curve inwards has angular corners.
While this effect remains a topic of research, we point out how
one may---absent knowledge of the correct result---address the
associated uncertainty as to which is the more accurate result.

\begin{figure}[t]
  \centering
  \includegraphics*[width=0.48\textwidth]{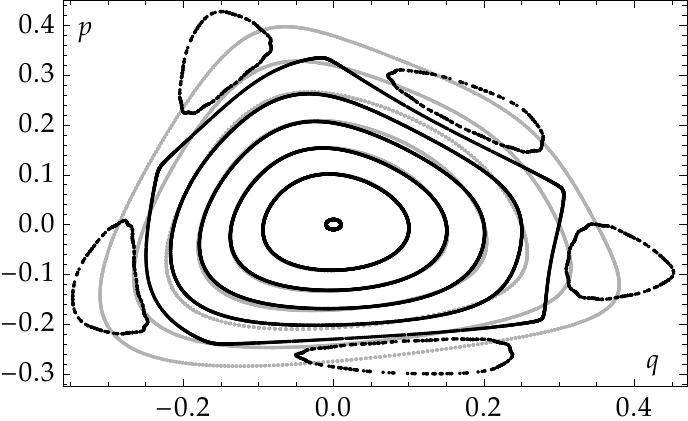}
  \hspace{3pt}
  \includegraphics*[width=0.48\textwidth]{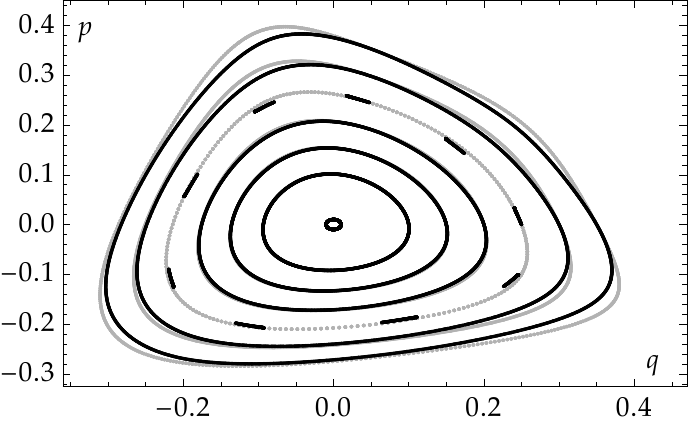}
  \caption{Phase-space portrait, in the case $\theta/2\pi = 0.22$,
    resulting from applying the map
    $\Mcr = \Map{R}\Ncr$
    repeatedly (2000~times) to the seven initial conditions
    $(q, p) = (0.01, 0)$, $(0.1, 0)$,     $(0.15, 0)$, $(0.2, 0)$,
             $(0.25, 0)$, $(0.3, 0)$, and $(0.35, 0)$
    to find their orbits.
    Light gray curves in the background indicate the exact result.
    In the left-hand graphic, the jolt maps are applied in numerical order,
    \ie~$\set{0,1,2,3,4}$. In the right-hand graphic, the jolt maps are
    instead applied in the order $\set{3,4,0,1,2}$.}
  \label{fig:iter.C.5j}
\end{figure}

Suppose we rewrite the map $\Map{RN}$ in the form
\begin{equation}
  \Map{RN} = \Map{R} \Map[1/2]{N} \Map[1/2]{N}.
\end{equation}
The reduced nonlinear content of the square root, $\Map[1/2]{N}$,
will make a Cremona symplectification of this map more accurate.
In words, this process means we
(i)~scale the Lie generator $\lieop{qp^2}$ by $1/2$;
(ii)~split $\Map[1/2]{N}$, as we did $\Map{N}$ in \eqref{N.cr2},
     to obtain  $\Nrtcr = (\Ncr)^{1/2}$; and
(iii)~square this map to obtain an approximation to \Map{M},
which we call $\Mrtcr$, in the form
\begin{equation}
  \Mrtcr = \Map{R}\Nrtcr\Nrtcr.
\end{equation}
(Note that in our example we now must apply ten successive jolt maps.)
This process, which we call the \emph{root trick}, can also improve
results obtained by use of generating functions%
  ~\bibnote[LM:S.34.3.5]{\emph{LM}, Section~34.3.5}.
The general technique of \emph{scaling, splitting, and squaring}
has application to a broad range of problems in computational physics%
  ~\bibnote[LM:S.4.1]{\emph{LM}, Sections~4.1 and 10.8}.

The left-hand graphic in \Fref{iter.C.5j.sqrt} shows the result of
applying the root trick to the ordering $\set{3,4,0,1,2}$.
Note the improved accuracy as compared to the right-hand graphic
in \Fref{iter.C.5j}.
Not shown is the result of applying the root trick to the ordering
$\set{0,1,2,3,4}$, which exhibits greatly improved accuracy.
In particular, gone are the offensive islands and angular corners
present in the left-hand graphic of \Fref{iter.C.5j}.


\begin{figure}[t]
  \centering
  \includegraphics*[width=0.48\textwidth]{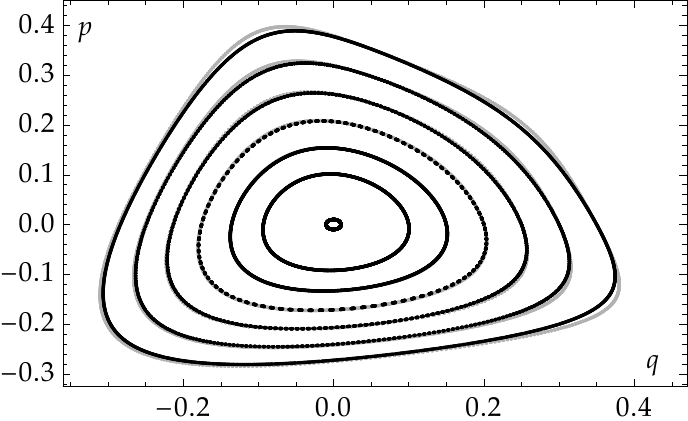}
  \hspace{3pt}
  \includegraphics*[width=0.48\textwidth]{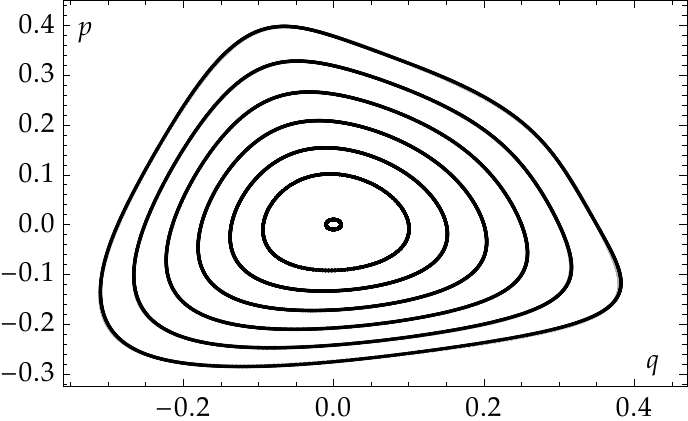}
  \caption{Cremona symplectification using five-plus-five jolts with
    the root trick.
    The left-hand graphic shows the result of applying the root trick
    to the map used for the right-hand graphic in \Fref{iter.C.5j}.
    And the right-hand graphic here shows the result of applying
    the \emph{symmetrized} root trick to that same map.}
  \label{fig:iter.C.5j.sqrt}
\end{figure}

Moreover, we can improve accuracy still further by \emph{symmetrizing} the factors:  By this we mean that one obtains an improved result by
approximating \Map{N} as
\begin{equation}
  \label{eq:N.rtscr}
  \Nrtscr = \Nrtcr \Nrtcr_\text{rev}
\end{equation}
where the subscript `rev' means that the given map's factors
should be applied in the \emph{reverse} order.
It can be shown that this symmetrization of the nonlinear factors
leads to an automatic reduction of many spurious higher-order terms
including the \emph{canceling} of \emph{all} $f_4$ terms.
(This is the same desirable feature found for the use of
a Poincar\'e generating function described in the previous section.)
The full map $\cal{M}$ now has the approximation
\begin{equation}
  \label{eq:M.rtscr}
  {\cal{M}}^{\rm{rtscr}} = \Map{R}{\cal{N}}^{\rm{rtscr}}.
\end{equation}

The right-hand graphic in \Fref{iter.C.5j.sqrt} shows the result of using
$\Mrtscr$, with the ordering $\set{3,4,0,1,2}$ for $\Nrtcr$,
to track particle trajectories.
Evidently, there is now near perfect agreement with exact results,
agreement that is comparable to that found with the use of
a Poincar\'{e} generating function.

We have illustrated how to find, over a substantial region of phase space,
a Cremona approximation for a nonlinear map (in one degree of freedom)
having the form ${\cal{N}}=\exp(:f_3:)$.
In an analogous (but substantially more complicated) manner,
with the use of more jolt maps, one may find,
in two and three degrees of freedom,
suitable Cremona approximations to maps of the form \eqref{map.DF.tr}.
Key to the construction of jolt maps for an $\Map{N}$ having $f_{>2}$
in its factored product form, and acting on phase spaces corresponding
to two or three degrees of freedom, is \emph{an optimal choice of the
linear maps $\Map{L}_j$} used in constructing the jolt maps.
It is known that an optimal choice of linear maps is related to the
construction of suitable \emph{cubature} formulas for various manifolds.%
  \footnote{Cubature formulas are higher-dimensional analogs
            of quadrature formulas.}
How to do so in the case of one degree of freedom is well understood
for all $f_{>2}$;
and the case of two degrees of freedom is reasonably well understood
for all $f_{<15}$.
Much work remains for the case of three degrees of freedom.
In particular, one would like to have, for that case, cubature formulas
for the manifold $SU(3)/SO(3)$.
Using this approach, a suitable set of 108 $\Map{L}_j$ has been found
for all $f_{<7}$%
  ~\bibnote[TMCA:PtII]{See \emph{TMCA}, Part~II.
     Also see 
     D.\,T. Abell, E.~McIntosh, and F.~Schmidt,
     ``Fast symplectic map tracking for the CERN Large Hadron Collider,''
     \emph{Phys. Rev. ST Accel. Beams}, vol.~6, 064001, June 2003.%
   }.

%% file: Conclude.tex

\section{Concluding Discussion}
\label{sec:conclude}

By design a storage ring has a (one-time-around) \emph{closed} orbit.
Even if the design is not perfectly executed, there is a
fixed-point theorem to the effect that there is still a closed orbit
that is near the design closed orbit.
Consider the passage of particles near the closed orbit through
individual beam-line elements or through collections of successive elements
(called \emph{lumps}) or once around the entire ring.
Each such passage is described by a symplectic map
whose jet is computable/known to some order $(maxm-1)$.
Each of these jets can be symplectified to produce symplectic maps
that can be used to propagate particles around the ring by letting them
act in succession thereby producing in effect
a \emph{net} one-turn symplectic map.
This operation is called \emph{tracking}.
The slowest, but presumably most accurate, procedure
would be to track element-by-element.
A faster procedure would be to track lump-by-lump.
Its accuracy could be checked by comparing its result to
element-by-element tracking results.
Even faster and more daring would be to perform full
turn-by-turn tracking using the symplectic map produced by
symplectifying the jet for the full one-turn map.
Whatever method is selected, it can be applied repeatedly
a large number of times to simulate the effect of a large number of turns
while being exactly symplectic (to machine precision)
and having accuracy through order $(maxm-1)$.
Since evaluation of the action of these symplectified maps
(obtained either by generating function or Cremona symplectification)
on phase space is fairly fast, it is possible to track in these ways
for a relatively large number of turns with relatively modest use
of computer time.

For example, let us consider the case of the LHC, for which a particle
passes through approximately 19,000 elements per turn.
One approach that has been used in the past is to track particles
element by element using the approximation given by
\eqref{approx} and \eqref{K.approx}.
That is, all fringe-field effects are neglected.
Moreover, the maps for drift spaces, bending magnets (dipoles),
and focusing/defocusing magnets (quadrupoles)
are approximated by linear (matrix) maps.
Finally, the nonlinear effects of higher multipole magnets
(sextupoles, octupoles, etc.), as well as multipole errors
in dipoles and quadrupoles, are simply treated as kicks.
This approach is often referred to as \emph{direct}/``\emph{exact}''
tracking, although what it actually does is equivalent to implementing
a relatively crude, but exactly symplectic, one-turn map.

Since generating function or Cremona tracking begins with symplectic jets,
and these jets can in principle be computed for \emph{realistic}
electromagnetic fields, generating function or Cremona tracking can in
principle be expected to give more accurate results for realistic machines.
However, in order to assure the Accelerator Physics community of their
reliability, generating function or Cremona tracking should also be able
to reproduce the results of direct tracking.
That is, based on the assumptions made for direct tracking,
relatively crude but symplectic jets can be computed for each
beam line element, and these jets can be concatenated to form
symplectic jets for lumps or one-turn maps.
These jets can then be symplectified using generating function or
Cremona methods, and their tracking results compared with those obtained
by direct tracking.
Preliminary studies/comparisons of this kind for the LHC,
prior to its completed construction, were carried out for various
nonlinear imperfection models.
They show that, even for relatively large betatron amplitudes
where nonlinear effects are expected to be important,
there is good agreement in dynamic aperture
(phase-space region of long-term storage)
between direct and one-turn Cremona map tracking results
using jets containing generators $f_{<8}$.
Moreover, in these studies Cremona tracking is approximately
20 times faster than direct tracking.
And, if one wishes to simulate orbits with smaller betatron amplitudes,
amplitudes associated with normal LHC operation, then use of generators
with $f_{<6}$ appears to be adequate, in which case Cremona tracking
is about 60 times faster than direct tracking.
Finally, the same Cremona map tracking speeds can be achieved
to produce accurate results for realistic machines.
Generating function and Cremona symplectification methods
are therefore worth further study, development, and implementation.

%% file: Ackn.tex

\section*{Acknowledgements}
\addcontentsline{toc}{section}{Acknowledgements}

We are grateful to the U.\,S. Department of Energy Office of Science
for research support over the years on the use of Map and Lie-Algebraic
methods in Accelerator Physics.
In addition, we thank RadiaSoft LLC for partial support provided
to one of us (DTA) during the preparation of this paper.